\documentclass[usenatbib]{mn2e}
\usepackage{graphicx, txfonts, natbib}
\usepackage{epsf}
\usepackage{amssymb}
\usepackage{epstopdf}
\usepackage{longtable}
\usepackage{lscape}
\usepackage{upgreek}
\usepackage{subfigure}

\newcommand{\msun}{\mbox{M$_{\odot}$}}

\newcommand{\kms}{\mbox{$\rm{km}\,s^{-1}$}}

\newcommand{\nick}{\mbox{$^{56}$Ni}}
\newcommand{\cob}{\mbox{$^{56}$Co}}

\DeclareMathAlphabet{\mathsc}{OT1}{cmr}{m}{sc}
\def\testbx{bx}%
\DeclareRobustCommand{\ion}[2]{%
\relax\ifmmode
\ifx\testbx\f@series
{\mathbf{#1\,\mathsc{#2}}}\else
{\mathrm{#1\,\mathsc{#2}}}\fi
\else\textup{#1\,{\mdseries\textsc{#2}}}%
\fi}
\newcommand{\ha} {\mbox{H$\alpha$}}

\newcommand{\Nai}{\ion{Na}{i}}
\newcommand{\Feii} {\ion{Fe}{ii}}
\newcommand{\Niii} {\ion{Ni}{ii}}
\newcommand{\Coii} {\ion{Co}{ii}}
\newcommand{\FeiiF}{[\ion{Fe}{ii}]}
\newcommand{\FeiF}{[\ion{Fe}{i}]}
\newcommand{\Fei}{\ion{Fe}{i}}
\newcommand{\Caii} {[\ion{Ca}{ii}]}
\newcommand{\CaiinF} {\ion{Ca}{ii}}
\newcommand{\Ci} {\ion{C}{i}}
\newcommand{\CiF} {[\ion{C}{i}]}

\newcommand{\Hei} {\ion{He}{i}}

\newcommand{\Oi} {[\ion{O}{i}]}
\newcommand{\Oinir} {\ion{O}{i}}

\newcommand{\Mgii} {\ion{Mg}{ii}} %not forbidden for 2800\AA\ line

\begin{document}

\title[Nebular spectral analysis of IIP SNe]
  {Constraining the physical properties of Type II-P supernovae using nebular phase spectra}

\author[K. Maguire et al.]
	  {K.~Maguire$^{1,2} \thanks{E-mail: kate.maguire@astro.ox.ac.uk}$, A.~Jerkstrand$^3$, S.~J.~Smartt$^2$, C.~Fransson$^3$, A.~Pastorello$^4$, S.~Benetti$^4$, 
 \newauthor  S.~Valenti$^4$,  F.~Bufano$^5$, G.~Leloudas$^6$\\
   $^{1}$Department of Physics (Astrophysics), University of Oxford, DWB, Keble Road, Oxford OX1 3RH, UK\\
    $^{2}$Astrophysics Research Centre, School of Maths and Physics, Queen's University Belfast, Belfast BT7 1NN, UK\\
    $^3$Department of Astronomy, The Oskar Klein Centre, Stockholm University, SE-106 91 Stockholm, Sweden\\
    $^4$INAF Osservatorio Astronomico di Padova, Vicolo dell'Osservatorio 5, 35122 Padova, Italy\\
   $^5$INAF Osservatorio Astronomico di Catania, Via S.~Sofia 78, 95123 Catania, Italy \\
   $^6$Dark Cosmology Centre, Niels Bohr Institute, University of Copenhagen, Juliane Maries Vej 30, 2100, Copenhagen, Denmark   }

%\date{}                                           % Activate to display a given date or no date
\maketitle
\begin{abstract}
We present a study of the nebular phase spectra of a sample of Type II-Plateau
supernovae with identified progenitors or restrictive limits. The evolution of line fluxes, shapes,
and velocities are compared within the sample, and interpreted by the use
of a spectral synthesis code. The small diversity within the dataset
can be explained by strong mixing occurring during the explosion, and by 
recognising that most lines have significant contributions from primordial metals in the H 
envelope, which dominates the total ejecta mass in these type of objects.
In particular, when using the \Oi\ 6300, 6364 \AA\ doublet for estimating the core
mass of the star, care has to be taken to account for emission from primordial O
in the envelope. Finally, a correlation between the \ha\ line
width and the mass of \nick\ is presented, suggesting that higher energy explosions 
are associated with higher \nick\ production.

\end{abstract}


\section{Introduction}
\label{intro}

Supernovae (SNe) are responsible for the production of most of the heavy elements that we observe in the Universe today. In a core-collapse SNe (CC-SN), this nucleosynthesis occurs both by hydrostatic burning during the star's evolution and by explosive burning behind the shock after collapse. \cite{woo95} analysed the yields of elements that are synthesised in massive stars and concluded that stars of 15-25 \msun\ are the typical stars responsible for producing heavy elements. However, the initial mass function shows that low mass stars are more common and therefore, depending on stellar yields, stars in the range 8--15 \msun\ could also make a non-negligible contribution to the chemical enrichment of the Universe.

Type II-Plateau (IIP) SNe are a subset of CC-SNe, which result from the core-collapse of massive H-rich stars, and therefore display prominent H features in their spectra, along with an extended plateau of nearly constant luminosity in their light curves. They are the most common SN type making up $\sim$40 per cent of all SNe per unit volume \citep{sma09}. Stellar evolutionary models predict that stars in the range 8--30 \msun\ should explode as red supergiants \citep{heg03, lim03, eld04, hir04}. However, the progenitors of some IIP SNe have been identified, and apart from  a few rare cases of blue supergiant progenitors \citep{wal87, pas05, kle11}, they have been found to be red supergiants in a lower than expected mass range of 8--17 \msun\ \citep{sma09}. \cite{utr08, utr09} have modelled the light curves and spectral evolution of IIP SNe (some with progenitor identifications) and find larger masses of the exploding stars than those from the direct detections of the progenitors.  This discrepancy between the progenitors obtained from direct imaging and those from modelling has also been analysed and discussed in \cite{sma09} and \cite{mag10} using a sample of nearby IIP SNe. One way of studying this discrepancy is to investigate and compare the nucleosynthetic yields from IIP SNe with identified progenitors. 

We do this by studying a sample of IIP SNe at epochs when we can see emission from newly synthesised heavy elements. During the plateau phase of the light curve, which generally lasts for $\sim$80--120 d, the light curve is powered by{remission of the energy deposited by the shock wave, including H recombinations. At these times, the H envelope is optically thick and we can not see emission from the metal core. However, after the H has recombined, the envelope becomes transparent and the core becomes visible. The light curve then enters the radioactive decay phase, which is powered by the decay of \cob\ to $^{56}$Fe. Radiative transfer effects are still significant in this phase, but decrease with time as the line optical depths fall. On the other hand, after approximately two years, the SNe generally become too faint to obtain good signal-to-noise spectra. Therefore, the optimal window for studying and interpreting the nucleosynthesis that has occurred is between one and two years post explosion.

Very few IIP SNe have been studied in detail during the nebular phase and few late-time nebular spectra of IIP SNe exist, with the notable exception of SN 1987A, which due to its proximity was very well monitored. This lack of late-time spectral data is partly due to the long exposure times that are required to obtain sufficient signal-to-noise. The best observed IIP SNe at the necessary late-times are SN 1987A \citep[e.g.,][]{phi90, bou93, mei93,wood93}, SN 1990E \citep{ben94}, SN 1999em \citep{leo02, elm03a}, SN 2004et \citep{sah06, mag10}, SN 2007it \citep{and11}, and SN 2007od \citep{and10, ins11}.

It is generally assumed that the emission lines of specific elements are emitted from regions containing predominantly that particular element, i.e.~that the O zone (formed in the late stages of pre-SN evolution or during the SN explosion) would be responsible for the production of the O emission lines. However, the final H envelope mass for an 8--12 \msun\ star is $\sim$6--8 \msun\ \citep[e.g.][]{hir04}, much larger than the mass of synthesised metals of $\sim$0.5 \msun\ and it is not clear if emission by primordial metals in the envelope may dominate that of synthesised ones. This has been shown by \cite{li93} to be the case in SN 1987A for the \Caii\ 7291, 7323 \AA\ and the \CaiinF\ 8600 \AA\ triplet, which are produced by a primordial abundance of Ca in the inner (velocity $<$ 2500 km s$^{-1}$) region of the H and He envelope and not by newly synthesised Ca. Ca emission lines are very efficient cooling lines, and so even small quantities of primordial Ca can result in prominent emission lines in the late-time spectra \citep{fra89}. To obtain the observed flux of the Ca lines, \cite{li93} needed only a Ca mass of 2$\times10^{-4}$ M$_{\odot}$ in $\sim$5 \msun\ of H in the inner envelope of SN 1987A, a factor of 10 times lower than that typically predicted for newly synthesised Ca in a SN 1987A-type progenitor star.  

Using models that included realistic mixing of the nuclear burning zones, \cite{koz98} showed that for SN 1987A, the \CaiinF\ 8600 \AA\ triplet was at all times formed by emission from primordial Ca in the envelope, whereas the \Caii\ 7291, 7323 \AA\ lines were formed by both synthesised and primordial Ca until $\sim$400 d, but mainly by primordial Ca after that. This does not imply that there is no newly synthesised Ca but instead that if it is present at late times, that the zones in which it is formed are too small and it cannot capture enough of the radioactive energy to create any strong emission, unless it is mixed with a greater mass of other elements such as H and He \citep{li93}. 

Therefore, to calculate the relative emission strengths from primordial and synthesised material, the deposited energy in the different zones, and the dominant processes responsible for the production of these lines must be modelled using a spectral synthesis code. Then the synthesised mass of O, Ca, and Fe can be estimated by first subtracting the emission contribution from primordial material in the envelope.
Previous studies of samples of CC-SN nebular phase spectra have mainly focused on stripped envelope SNe (Type Ib (lacking H) and Type Ic (lacking H and He)) \citep[e.g.][]{mae08, mod08, tau09, mil10}, and have studied the \Oi\ 6300, 6364 \AA\ profiles to look for possible signs of ejecta asphericity in the line profile shapes and how they evolve with time.  \cite{elm11} have studied a sample of  CC-SNe including IIP SNe as well Type Ib/c SNe, focusing again on an analysis of the \Oi\ 6300, 6364 \AA\ doublet. Using the overall luminosity of the \Oi\ doublet and the ratio of the two components, they have estimated the synthesised O mass, assuming all the O luminosity is produced by freshly made O and not from primordial O in the envelope. We will in this paper show that this is a questionable assumption.

In this paper, we study a sample of 25 late-time nebular phase spectra of nine IIP SNe with identified progenitors to investigate how their spectral features change as a function of time, and how the flux, profile shape and velocity of the features vary among SNe. This can then be linked to the material distribution and physical conditions in the SNe. We will apply a spectral synthesis model to the nebular phase spectra to study line flux ratios and interpret their evolution with time as well as their internal differences. This paper is complemented by Jerkstrand et al. (in prep.), which derives detailed nucleosynthesis results for one of the SNe in the sample, SN 2004et, by a complete analysis of the spectral evolution from the ultra-violet to far-infrared. 

\section{Supernova sample of nebular spectra}
\label{SN_sample}

As mentioned in Section \ref{intro}, the current sample of IIP SNe with multi-epoch optical spectra at late times is limited to a small number of objects. In particular we wanted to focus on SNe with information available about their progenitor stars and \nick\ mass estimates. To increase this sample, observations were obtained of four IIP SNe at the 3.6-m New Technology Telescope (NTT) with the ESO Faint Object Spectrograph and Camera version 2 (EFOSC2) in low resolution spectroscopy mode and at the 8.2-m Very Large Telescope (VLT) with FOcal Reducer and low dispersion Spectrograph 2 (FORS2). The SNe observed were SNe 2007aa, 2008bk, 2008cn, and 2009N at epochs spanning $\sim$360--550 d post-explosion. 
In addition to the new data set, the literature was searched for late-time nebular spectra of IIP SNe with progenitor information (either a progenitor has been identified in pre-explosion images or an upper mass limit has been set) and data from the following SNe were included, SNe 1987A, 1999em, 2003gd, 2004et, and 2005cs. The complete sample of spectra analysed here is detailed in Table \ref{sn_sample}. All the spectra have been corrected for Galactic and host galaxy extinction using the values given in Table \ref{sn_sample2} and they have also been corrected to the rest-frame of their respective host galaxies. The resolution of the spectra that are quoted were either taken from the literature or estimated from the full width at half maximum (FWHM) of the night sky lines. 

The \nick\ masses and zero-age main sequence masses of the SN progenitors, where available, are given in Table \ref{sn_sample2}. The properties of the sample range from low-luminosity, low \nick\ mass events such as SNe 2005cs, 2008bk, and 2009N to higher-luminosity, higher \nick\ mass events such as SNe 1987A, 1999em, and 2004et. However, the range in main-sequence masses listed in Table \ref{sn_sample2} does not show as large a spread between over and under luminous events, with the masses clustering at the lower end of the range expected from stellar evolutionary theory (8--30 \msun). We will investigate here if there are any measurable differences between the spectral properties of the IIP SNe in the sample and then determine how these differences might relate to their progenitor and explosion properties.

\begin{table*}
%\begin{flushleft}
\caption{Spectral observations of the sample of IIP SNe studied, including epoch, wavelength range, resolution and source of the spectrum. The SNe for which new data are presented in this paper are noted as VLT+FORS2 and NTT+EFOSC2 as well as the grism with which they were taken, as discussed in Section \ref{SN_sample}.}
%\end{flushleft}
\smallskip
 \label{sn_sample} 
\begin{tabular}{lccclccccccccccc}
\hline
\hline
 SN	 	&	Epoch & $\lambda$ Range&Resolution& Source \\ % Explosion&
& (d)& (\AA)  & (\AA)   \\ %epoch (JD)&
\hline
1987A &338&3110--10930&16&\cite{phi90} \\ 
&345&3100-10610&16&\cite{phi90}\\
&374&5160--10690&16&\cite{phi90}\\
&398&3200--10220&16&\cite{phi90}\\
&407&3020--7780&16&\cite{phi90}\\
&422&3070--10740&16&\cite{phi90}\\
&497&3000-10990&16&\cite{phi90}\\
&513&3030--10970&16&\cite{phi90}\\
&525&3000--10990&16&\cite{phi90}\\
&529&4000--10350&16&Asiago Supernova Archive\\
&586&3230--7500&16&\cite{phi90}\\
&598& 3830--10170&16&Asiago Supernova Archive\\
1999em &312 &3350--10620 &21& \cite{elm03a}\\ %2451475.6&
&319&3410--7770 &21& \cite{elm03a}\\
&391&3500--9110&10& \cite{elm03a}\\
&465&3360--10620 &20& \cite{elm03a}\\
&510&3560--9240&10& \cite{elm03a}\\
&642&4200--9940 &28& \cite{elm03a}\\
2003gd  & 493 & 4200--9600 &13&\cite{hen05} \\ %2452717&
2004et&284&4000--7820&36&\cite{mag10}  \\ %2453270.5&
&301 & 4000-8450&\phantom{0}7&\cite{sah06}\\
&314&4000-8450&\phantom{0}7&\cite{sah06}\\
&341   &3400--8020 &13 &\cite{mag10}\\ 
&384   & 3500--7750&24&\cite{mag10}\\
&401&4000-8450&\phantom{0}7&\cite{sah06}\\
&408   & 4000--7730&25 &\cite{mag10}\\
&427  & 4000--8450&\phantom{0}7&\cite{sah06}\\
&464   & 4000--8450 &\phantom{0}7&\cite{sah06}\\
2005cs &333   & 5050--10470&18 & \cite{pas09}\\ % 2453549.0&
2007aa &376&4300--9600&10&VLT+FORS2+300V\\ %2454131.0 &
2008bk &524&3500--9500&14&NTT+EFOSC2+gr\#11$^a$\\ %2454532.5&
& 547  &   3800--9500                      &10&VLT+FORS2+300V    \\
2008cn&358&3500--9500&10&VLT+FORS2+300V\\ %2454598&
2009N&366&6000-10100&14&NTT+EFOSC2+gr\#16$^a$\\ %2454855&
&406&3500-9600&10&VLT+FORS2+300V\\
\hline
\end{tabular}
\medskip\\
$^a$The resolution of the gr\#16 and gr\#14 gratings is very similar but gr\#16 is centred at a longer wavelength than the gr\#11.
 \end{table*}

\begin{table*}
\caption{Some of the derived properties for the sample of IIP SNe analysed in this paper including their extinction, distance, \nick\ mass and zero-age main sequence (ZAM) mass.}
\smallskip
 \label{sn_sample2} 
\begin{tabular}{llclllcccccccccc}
\hline
\hline
 SN&E(B--V)&Dist.	 &Ni$^{56}$ mass	&	ZAM&Source \\ % Explosion&
&&(Mpc)& (\msun)&(\msun)& \\ %epoch (JD)&
\hline
1987A &0.19& 49.9$\times 10^{-3}$ &0.075$\pm$0.005&14--20&1, 2\\ 
1999em &0.06&11.7$\pm$1.0&0.042& $<$ 15&3\\ 
2003gd  &0.14&9.3$\pm$1.8&0.016& 7$_{-2}^{+6}$&4 \\ 
2004et  &0.41& 5.9$\pm$0.4&0.064\phantom{0}$\pm$0.04  &8$_{-1}^{+5}$ &5, 6, 7  \\ 
2005cs &0.05&8.4$\pm$1.0&0.003& 8$\pm$2& 8, 9, 10\\ 
2007aa&0.014&20.5$\pm$2.6&--&$<$ 12&11, 12\\ 
2008bk&0.3&3.9$\pm$0.4  &0.004&9$_{-1}^{+4}$&13, 14\\ 
2008cn&0.35& 33.3$\pm$0.2 &--&13$\pm$2&15, 16\\ 
2009N &0.15& 12.6$\pm$0.9&0.009$\pm$0.004&$<$16& 16, 17 \\ 
\hline
\end{tabular}
   \medskip \\
REFERENCES -- (1) \cite{phi90}; (2) \cite{sma09b}; (3) \cite{elm03a} (4) \cite{hen05}; (5) \cite{sah06}; (6) \cite{mag10};  (7) \cite{cro11}; (8) \cite{mau05}  (9) \cite{li06} (10) \cite{pas09}; (11) \cite{sma09}; (12) \cite{cho10}; (13) \cite{mat08}; (14) \cite{van10}; (15) \cite{eli09}; (16) Fraser et al.~(in prep.); (17) Tak\'ats et al.~(in prep.)
\end{table*}

\section{Description of radiative transfer model} \label{model_setup}

Full details of the spectral synthesis model used to study the nebular phase spectra can be found in \cite{jer10} and Jerkstrand et al. (in prep.). Here we describe only the basic setup of the model and its important input parameters. The model uses as an input the 12 \msun\ explosion model of a solar metallicity progenitor taken from \cite{woo07}. A 12 \msun\ model is chosen because this mass is typical of the progenitor mass estimates obtained from pre-explosion images of the sites of IIP SNe and stellar evolutionary models \cite[][and references therein]{sma09}. The late stellar evolution and explosion of stars less massive than 12 \msun\ involve significant uncertainties, and few if any useful models are available. In Jerkstrand et al. (in prep.) we discuss models for a wider set of masses. 

The ejecta is divided into a spherically symmetric core and a spherically symmetric envelope, with the core containing the zones of the heavier elements (Fe/He, Si/S, O/Si/S, O/Ne/Mg, O/C), along with 15 per cent of the mass of H and 60 per cent of the mass of He mixed in from the envelope. The zones making up the core region are broken into clumps (100 for each zone), with the clumps of each zone having the same density and being randomly distributed throughout the core region.  This simulates macroscopic mixing in the core and removes the rigid shell-like structure that is known not to be an accurate representation of the ejecta structure post-explosion, as will be discussed in Section \ref{compsec}. We allocate the filling factors using the same method as in \cite{jer10}, assuming the same density for the synthesised O as derived for SN 1987A by \cite{li92}. At epochs in the range 100--1000 d post-explosion, the main source of luminosity is due to the radioactive decay of \cob\ and its mass can be constrained by measuring the bolometric luminosity during the radioactive tail phase of the SN. For this 12 \msun\ model, a \nick\ mass of 0.064 \msun\ and a core velocity of 1800 \kms\ were chosen to agree with the values obtained from observations of SN 2004et. The model computes the deposition of gamma rays and positrons, followed by iterative solutions to the equations of statistical and thermal equilibrium, as well as the radiation field. See \cite{jer10} for more details. We note that for low velocity/low \nick\ mass SNe such as SN 2008bk, the conditions for the steady state calculation assumed in the model may not be met at all times, and the model line fluxes are therefore of limited accuracy for these types of SNe.

\section{Line profile analysis} \label{compsec}

Standard stellar evolution models of massive stars \cite[e.g.][]{woo95,thi96,hir04} describe how the pre-explosion structure of the progenitor has an ordering of layers of elements from each nuclear burning stage, with the outermost layer being H and the innermost Fe.  In the absence of mixing, this onion-layer structure would hold for SNe after explosion and so if the emission of each particular element is formed in a different shell containing predominantly that element, then there would be a correlation between the elemental mass numbers and the width
of their emission lines. The lightest elements (H, He) would have broad lines with pronounced flat-tops, and the heaviest elements (Si, Si, Fe, Ni, Co) would have narrow lines. At the other extreme, if the elements were distributed uniformly throughout a sphere, then the line profiles would be parabolic in shape. 

Studies of the observational properties of SN 1987A have found very similar line widths for a wide range of elements with no flat-tops, which suggests extensive mixing of the ejecta \citep{fra87,mei89, mei93}. This is supported by the early detection of gamma rays and X-rays from SN 1987A \citep{dot87, sun87}, which would only occur if $^{56}$Co and $^{57}$Co were mixed out into the inner envelope in concentrated clumps \citep{pin88, kum89}. The detection of a high excitation line, \Hei\ 10830 \AA\ at times greater than 10 d \citep{eli88} could also only occur if it was reionised by gamma rays emitted in the He zones from outward mixed, decaying \cob\ \citep{gra88, fas98}.

\begin{figure*}
\includegraphics[width=18cm]{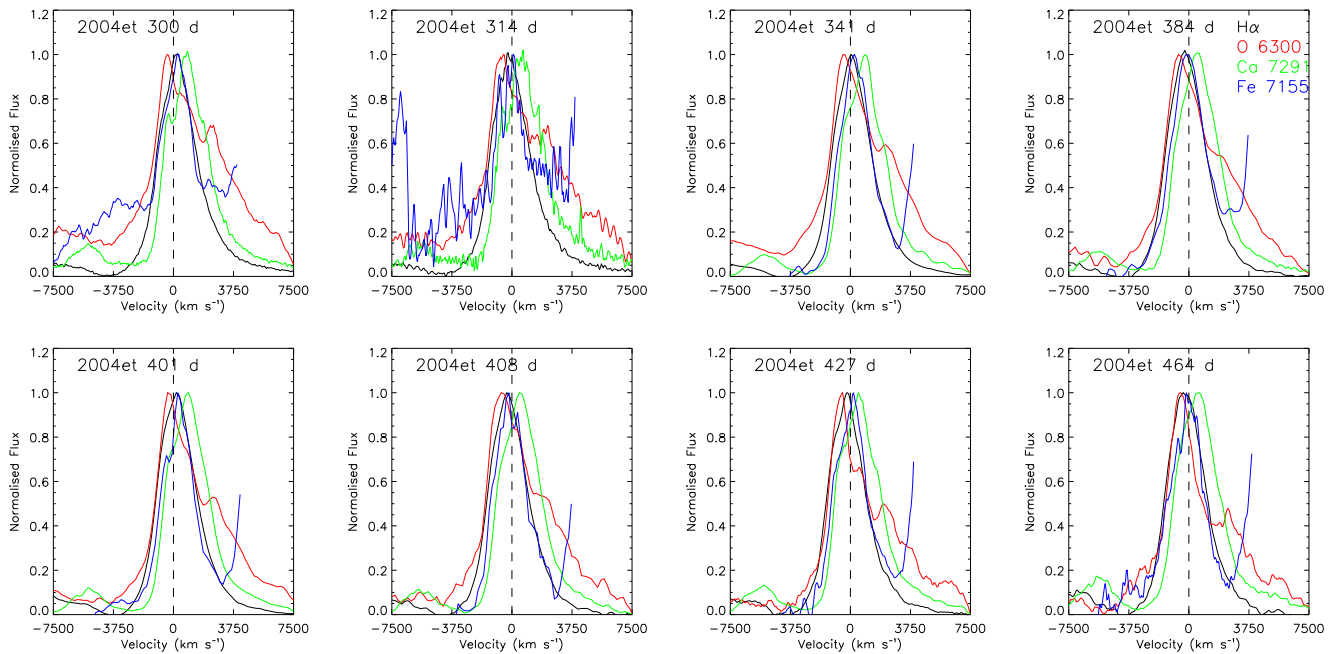}
\caption{Line profiles for SN 2004et between 300 and 464 d. The \Oi, \Caii\ and \FeiiF\ lines shown are doublets and have been centred on the blue component, as discussed in Section \ref{shapes}. The profile of \ha\ is shown in black, \Oi\ 6300, 6364 \AA\ in red, \Caii\ 7291, 7323 \AA\ in green, and the \FeiiF\ 7155, 7172 \AA\ doublet is shown in blue.}
\label{04etearlyprofile}
\end{figure*}

\subsection{Line profile shapes} \label{shapes}
Here we investigate the shapes of the profiles of the nebular phase emission lines in multi-epoch spectra for our sample of SNe, to study the distribution of elements in velocity space, and determine if signatures of mixing are also seen in the spectra of our sample. Figure \ref{04etearlyprofile} shows the velocity profiles of the prominent emission lines present in SN 2004et at 300--464 d, while Figure \ref{99emprofile} shows the velocity profiles for SN 1999em in the range 312--510 d, and Figure \ref{lowvelprofile} shows the profiles for low-luminosity SNe 2008bk (524, 547 d) and 2009N (366, 406 d). The lines shown are \ha, \Oi\ 6300, 6364 \AA, \Caii\ 7291, 7323 \AA, and \FeiiF\ 7155, 7172 \AA. Each of the lines shown has had a linear continuum subtracted and has been normalised to its peak value. The \Oi, \Caii, and \FeiiF\ lines are doublets and to compare the emission line shapes, the profiles have been centred on the rest wavelength of the blue components of the doublets. 

None of the emission lines of these SNe shown in Figures \ref{04etearlyprofile}, \ref{99emprofile} and  \ref{lowvelprofile} has a flat-topped profile at any epoch and the lines are instead roughly Gaussian in shape. However, the convolution of the emission lines with the point spread function (PSF) of the instrument could result in a roughly Gaussian shape being seen for an intrinsically flat-topped profile if the intrinsic width of the line is smaller or comparable to the resolution of the instrument. To investigate the effect, we have convolved a flat-topped profile (with the width of the \ha\ line) with a Gaussian (with the width of the instrumental resolution). However, before doing this, the observed FWHM (FWHM$_{\rm obs}$) must be corrected to account for the resolution of the instrument (FWHM$_{\rm instr}$) as measured from the FWHM of the night sky lines in each of the spectra to obtain the original width of the line before convolution with the telescope instrument. The corrected FWHM is given by,

\begin{equation}
\label{fwhm_corr}
\displaystyle{FWHM_{\rm corr} = \sqrt {FWHM_{\rm obs}^2 - FWHM_{\rm instr}^2}}
\end{equation}
Then for each SN in the sample, we convolved a flat-topped line profile (with the width of the \ha\ line, corrected for the instrumental resolution) with a Gaussian (with the width of the instrumental resolution) to determine if any of the line profiles were flat-topped. 

To demonstrate this, in Figure \ref{flattop}, a comparison between the convolution of the resolution of the instrument with both a flat-topped and Gaussian profile is shown for two different FWHM$_{\rm corr}$ values.  In the top panel of Figure \ref{flattop} using a FWHM$_{\rm corr}$ of 20 \AA, it is not possible to distinguish from the output between the two scenarios. However, in the bottom panel with a FWHM$_{\rm corr}$ value of 28 \AA, the result is different, with the flat-topped profile giving a broader, flatter-topped line profile. The FWHM at which it becomes difficult to distinguish between the intrinsic profile types for the quoted resolutions is a FWHM of $\sim$29 \AA\ for the NTT+EFOSC2 (resolution of 14 \AA) and a FWHM of $\sim$22 \AA\ for the VLT+FORS2 (resolution of 10 \AA) at the position of \ha\, assuming a signal-to-noise of greater than $\sim$20. 

We have found for all the SN spectra in our sample that it is possible to distinguish between these two scenarios and so the roughly Gaussian shaped profiles that are seen are found to be intrinsic to the SN emission, and this immediately suggests that mixing of the elements must occur in the SN ejecta to allow elements to be at zero velocity. Only for the spectrum of the narrow-lined SN 2008bk obtained at the NTT+EFOSC2 with a resolution of 14 \AA, was it found not to be possible to distinguish between the intrinsic Gaussian profile and a flat-topped one. However, another spectrum of SN 2008bk was obtained at the VLT+FORS2 with a resolution of 10 \AA, where it was possible to distinguish between these two scenarios and the line profile of \ha\ of SN 2008bk was found not to be intrinsically flat-topped.

In Figures \ref{04etearlyprofile}, \ref{99emprofile} and  \ref{lowvelprofile}, it should be noted that the profiles of the different lines for a particular SN appear to have similar line profile shapes, which again suggests that the different elements producing the emission lines have a similar spatial distribution in the SN ejecta.  In the early spectra of SN 2004et in Figure \ref{04etearlyprofile} and for SN 1999em in Fig. \ref{99emprofile}, it appears as if the blue wing of the \Oi\ 6300 \AA\ line extends to higher velocities than the other lines. However, this early blue wing of the \Oi\ 6300 \AA\ line disappears at later times and is most likely caused by a blend with another emission line. In particular, \cite{des10} showed that there are many weak Fe lines in this region that could be blending with the \Oi\ doublet \citep[see also][]{fra02}. If the H emission lines were being produced solely in the envelope and the other elements from a central region then it would be expected that the wings of the H profiles should extend to greater velocities, which does not appear to be the case. However, the velocities that can be measured from these emission lines are wavelength dependent and also must be corrected for the resolution of the instrument used to obtain the observations and therefore, a more quantitative analysis of this is presented in Section \ref{vel_dist}. 

An evolution with time in the line profile shapes could be expected if at early times the emission mainly comes from regions close to the \nick\ but then as the gamma rays propagate outward, emission from other regions is seen. However, no evolution is seen with time, which could suggest that \nick\ is not concentrated in the core of the SN but instead is distributed by mixing to regions farther out in the ejecta.

Asymmetry of some of the line profiles is seen in Figures \ref{04etearlyprofile}, \ref{99emprofile} and \ref{lowvelprofile}. The nebular phase \Oi\ 6300, 6364 \AA\ doublet of SN 2004et was investigated by \cite{utr09} for signs of possible asymmetry in the ejecta. They fitted three Gaussian components to each of the lines of the \Oi\ doublet and inferred a bipolar structure of the line-emitting gas in the inner layers of the SN envelope. They noted that this asymmetry refers to the line-emitting gas, which is not identical to the overall O distribution. An attenuation of the red wing (blueshift) of the line profiles of \ha\ and \Oi\ is seen for SN 2004et at $\sim$300 d and shown in \cite{mag10} to be coeval with other indicators of dust formation. Dust condensation within the metal-rich ejecta of SN 1987A was first noted by \cite{luc89} from the asymmetry of optical emission lines. Signatures of dust formation such as line attenuation were also seen for SN 1999em \citep{elm03a} at $\sim$500 d post explosion, and SN 2009N appears also to show a blueshifted profile for the \Oi\ 6300 \AA\ line at both the epochs shown in Figure \ref{lowvelprofile}, which is most likely due to dust formation in the ejecta. 

Other signs of dust formation include a decrease in luminosity at optical wavelengths accompanied by an increase at NIR wavelengths. Signatures of dust formation have been seen at mid-infrared wavelengths of IIP SNe to occur at similar epochs to the appearance of the blueshifted profiles for SNe 1987A \citep{mei93}, 2003gd \citep{sug06, mei07}, 2004dj \citep{mei11}, and 2004et \citep{kot09}. Therefore, the shapes of the emission lines along with other observational parameters, tell us not only about the velocity distribution of the ejecta but also about the onset of dust formation and any possible asymmetric distribution of elements in the ejecta. 

 \begin{figure}
\includegraphics[width=8.4cm]{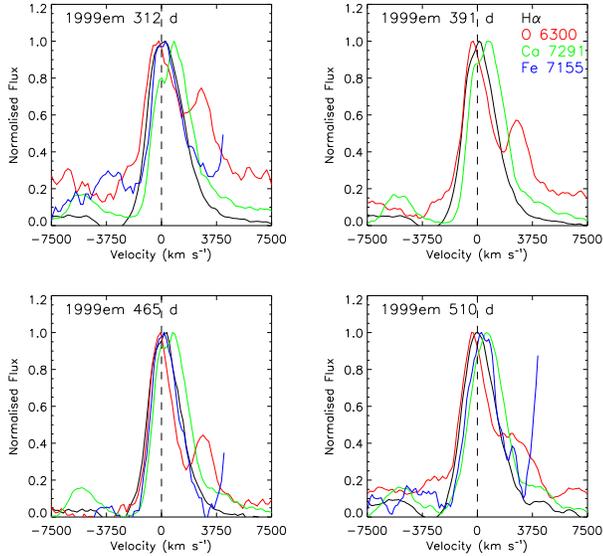}
\caption{Line profiles for SN 1999em between 312--510 d. The lines are as specified in Figure \protect \ref{04etearlyprofile}.}
\label{99emprofile}       
\end{figure}

\begin{figure}
\includegraphics[width=8.4cm]{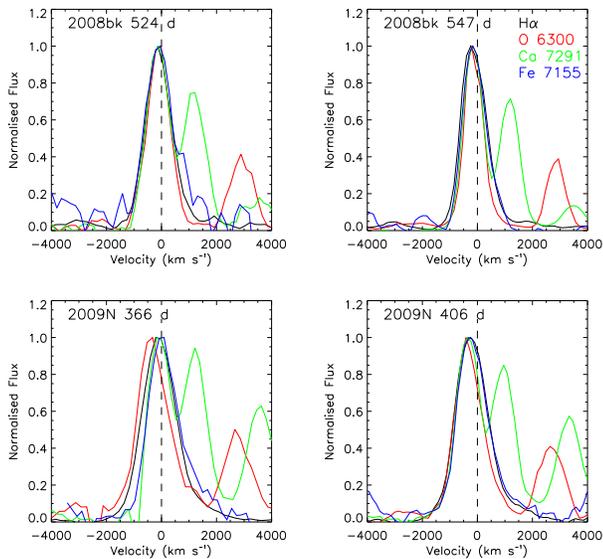}
\caption{Line profiles for two low luminosity SNe 2008bk and  2009N. The lines are as specified in Figure \protect \ref{04etearlyprofile}.}
\label{lowvelprofile}
\end{figure}

\begin{figure}
\includegraphics[width=9cm]{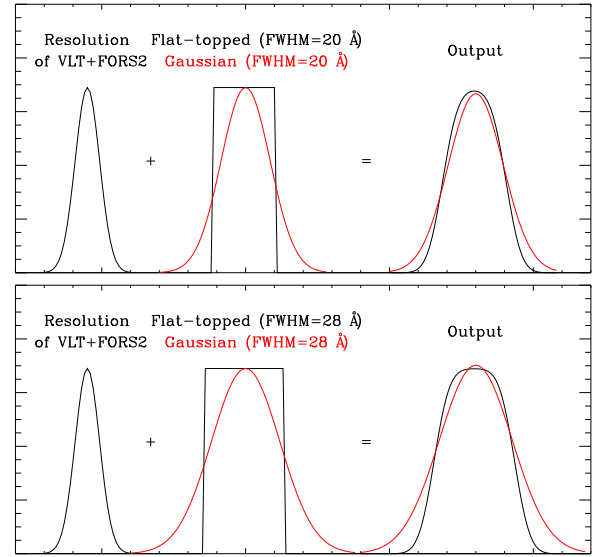}
\caption{Comparison of the convolution of the instrumental resolution obtained with the VLT+FORS2 of 10 \AA\ (measured from night sky lines) with both a flat-topped and a gaussian profile. In the top panel, a FWHM$_{\rm corr}$ of 20 \AA\ is used, which corresponds to a post-convolution value of $\sim$22 \AA\ and it can be seen that the output line profiles are very similar and it is not possible to distinguish between them. In the bottom panel, an input FWHM$_{\rm corr}$ of 28 \AA\ is used, which corresponds to a post-convolution value of $\sim$30 \AA\ and the output line profiles for the flat-topped and gaussian inputs are found to be different.}
\label{flattop}
\end{figure}

\subsection{Velocity distribution} \label{vel_dist}
To study the distribution of elements in the ejecta and velocity at which the bulk of the emission occurs, the FWHM of the prominent emission lines in the sample of SNe were measured by fitting Gaussian profiles to each of the lines. The FWHM used were corrected as discussed in Section \ref{shapes} for the resolution of the instrument used to obtain each spectra using Equation \ref{fwhm_corr}. From these corrected FWHM values, the half width at half maximum (HWHM) velocities can be obtained to determine the velocity to which the different elements are found in the ejecta with respect to the zero velocity of the line profile. 

The HWHM velocities of the sample of SNe studied here are shown in Figure \ref{vel_all}. The SN sample is split into three groups based on the average velocities of the emission lines. For the \Oi\ and \Caii\ doublets, the average value of the measurements of the two lines was used. Limits are shown for some of the emission lines when the individual components were not well enough resolved to fit Gaussians to each of the lines and were obtained by fitting a single Gaussian to the whole line profile. Therefore, these limits are very conservative upper limits for the widths of the individual lines. The outer wings of the line profiles could also be calculated from either the position at which the line flux goes to zero or the location of the first local minimum outward from the centre. The position of these outer wings would then tell us the velocity to which a certain element extends. However, the uncertainties in measuring these values are large due to the effects of blends of lines, contributions from the continuum and noisy spectra where the local minimum does not represent the true line minimum. Therefore, like \cite{des10}, we have decided to use the more reliable HWHM for calculating the velocities, since it is more easily measured and compared among SNe.

Extensive studies of the emission line profiles of SN 1987A have been performed in the past and as a consistency check, we have compared our measurements of SN 1987A to those of previous studies. \cite{mei93} studied 1--4 $\upmu$m spectroscopy of SN 1987A and found that the \Oinir\ 1.129 $\upmu$m velocities were lower than those of H, in agreement with Figure \ref{vel_all}. The \FeiiF\ and \Caii\ velocities of SN 1987A have only upper limits plotted, but the \Caii\ lines are consistent with having been produced in the inner regions of the H envelope as detailed in \cite{li93}. The \Oi\ velocities are also consistent with those summarised in \cite{li92}. \cite{li93b} found that the \Feii, \Coii, and \Niii\ lines of SN 1987A have FWHM of 3000-3500 km s$^{-1}$, which are consistent with the HWHM velocity limits of \FeiiF\ shown here. The measurements of SN 1987A are consistent with those of previous studies, which suggests that our method of fitting the Gaussian profiles is sound. 

\begin{figure*}
\includegraphics[width=18cm]{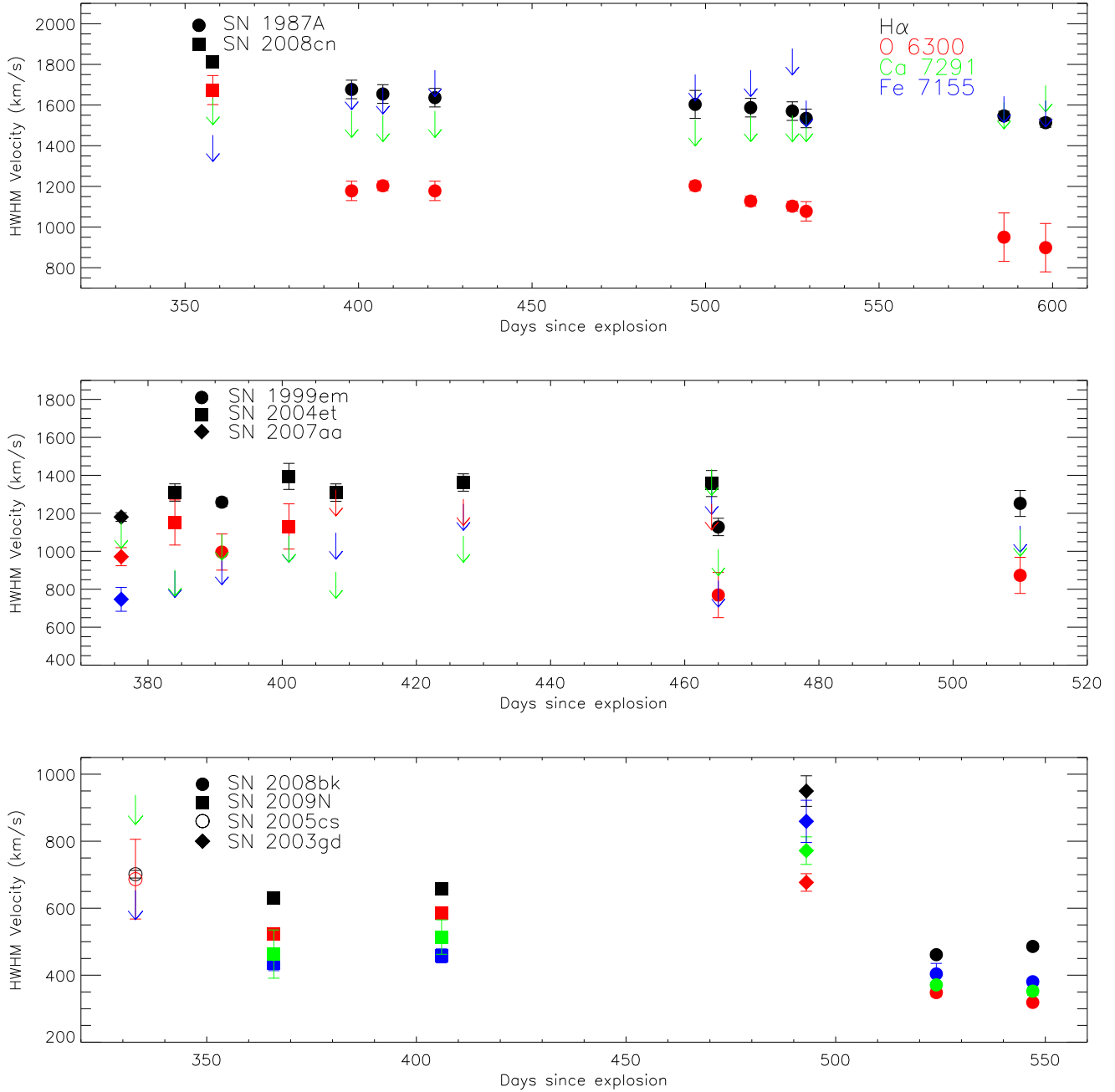}
\caption{The HWHM (half the FWHM) velocity of the emission lines of \ha\ (black), \Oi\ (red), \FeiiF\ (blue), and \Caii\ (green) of SNe 1987A and 2008cn (top panel), SNe 1999em, 2004et, and 2007aa (middle panel), and SNe 2008bk, 2009N, 2005cs, and 2003gd (bottom panel) are shown. The SNe are grouped based on the average velocities of the emission lines. These velocities have been corrected for the intrinsic resolution of the instrument used to obtain each spectrum.}
\label{vel_all}
\end{figure*}

In Figure \ref{vel_all}, there is a much larger difference between the \ha\ and \Oi\ velocity for SN 1987A than for the other SNe in the sample. This could suggest that the H envelope plays a role in contributing to the flux seen in the prominent emission lines for the other SNe, since the velocities for the different elements are relatively similar. The immediate physical interpretation of this is that a major fraction of the \Oi\ emission does not come from newly synthesised O, but from primordial O that is present in the H zones or from O clumps mixed to high velocities. For the low-luminosity, low velocity events (SNe 2005cs, 2008bk, and 2009N) shown in Figure \ref{vel_all}, actual measurements (instead of limits) of the expansion velocities were obtained for all the lines and it can be seen that \ha\ has higher velocities than the other lines, although the difference is not as great as in SN 1987A. 

In \cite{des10}, radiation-hydrodynamic simulations of IIP SNe were performed. They used a 1D model without mixing and suggested that their models of the chemical distribution of the ejecta in velocity space can be used to place strong constraints on the main-sequence masses of IIP SN progenitors. As shown above, the line profile shapes indicate that mixing is likely to have occurred 
in the ejecta of IIP SNe, and this argues against the assumption of chemical stratification 
in the SNe. While the \cite{des10} models are an informative 
first step in modelling nebular spectra to probe progenitor and explosion characteristics, 
the mixing of material in velocity space is likely an important parameter for future
exploration. \cite{des10} used the width  of the \Oi\ 6300, 6364 \AA\ doublet to estimate the velocity of the O-rich zones as it was shown for SN 1987A at $\sim$500 d by \cite{koz98} that the \Oi\ lines were formed in the O-rich zones. However, in Section \ref{model_results} we describe how a radiative transfer model can be used to determine the contribution to line emissions from different zones in the SN ejecta and it is shown for a 12 \msun\ model that the flux in the \Oi\ 6300, 6364 \AA\ doublet has approximately equal contributions from primordial O in the H-rich zones and synthesised O in the O-rich zones. However, as the mass of the progenitor increases, as in the case of SN 1987A, the contribution from the O core becomes dominant.

To investigate the effect of O emission from the envelope on the measured FWHM of the \Oi\ 6300, 6364 \AA\ doublet, we used the spectrum resulting from the model detailed in Section \ref{model_setup}. We compared the FWHM values obtained from the model spectrum with and without the contribution from the H zones and found that by excluding the contribution from the H zones, the FWHM of the \Oi\ 6300, 6364 \AA\ doublet in the model spectrum was reduced by $\sim$8 \AA, which corresponds to a HWHM velocity difference of $\sim$200 \kms. This would suggest that using the HWHM of the [O I] 6300, 6364 \AA\ lines 
as an indicator of the extent of the spatial distribution of O would result in small but measurable
overestimates of the velocities.  The relatively small reduction of $\sim$8 \AA\, when excluding the contribution from the H zones is due to the fact that while half of the \Oi\ flux is emitted by these zones, the main H zones contributing to the \Oi\ flux are those that are mixed into the core and therefore, have a lower velocity, than the average of all the H zones.

\section{Line flux analysis} \label{outputs}
\subsection{Model results} \label{model_results}

The synthesised and primordial masses from our 12 \msun\ model are given in Table \ref{masses} for some of the main elements, along with the ratio of primordial to synthesised mass for each of these elements. It can be seen that the synthesised masses are all larger than the primordial masses. However, when it comes to the observed spectrum it must be remembered that the flux in the emission lines will also depend on how efficient an emitter the element is per unit mass and how the energy is deposited in the ejecta. 

\begin{table}
\caption{Primordial and synthesised masses of the listed elements from the 12 \msun\ model used in this paper, along with the ratio of primordial to synthesised mass for each of the elements. The Fe mass is calculated at 400 d for a \nick\ mass of 0.064 \msun.}
\label{masses}
\begin{tabular}{ccccccccccccccccccccccccccccc }
\hline
\textbf{Element} &Synthesised&Primordial& Synthesised/Primordial \\
&Mass (\msun)&Mass (\msun)& Mass Ratio\\
\hline
O  & 3.08E-01 & 4.45E-02  &  6.9\\
      Si & 4.12E-02 & 7.12E-03  &  5.8\\
      S  & 2.63E-02 & 3.65E-03  &  7.2\\
      Ca & 2.40E-03 & 6.41E-04  &  3.7\\
      Fe & 6.5E-02 & 1.22E-02  &  5.4\\
\hline
\end{tabular}\\
\end{table}

The spectral synthesis model can be used to calculate the contributions to the luminosity of the prominent emission lines observed during the nebular phase from the various regions of the SN ejecta. In this section, we compare the output of the model to the observations in the range of 300--650 d. We make specific in-depth comparisons at 400 d, when most of the SNe in the sample have observed spectra. At this epoch, our model with a \nick\ mass of 0.064 \msun, gives a total deposited energy of 1.9 $\times$ 10$^{40}$ ergs, which is $\sim$75 per cent of the emitted energy ($\sim$25 per cent of the gamma rays escape). The main characteristics of the model at 400 d  are detailed in Table \ref{model_tab}, including the energy deposited in each of the zones as well as the channels through which this deposited energy is released (heating of the gas, ionisations and excitations) and the main cooling lines. In this table, F$_{i}$ is the fraction of the energy deposited into  zone $i$, $f_{exc}^{i}$, $f_{ion}^{i}$ and $f_{heat}^{i}$ are the fractions of this energy going into non-thermal excitations, ionizations and heating, respectively. Finally, $f_{\lambda}^{i}$ is the fraction of the cooling done by a given line in a given zone. 

It can be seen that approximately 2/3 of the total energy is deposited in the H zones, with the next highest amount of deposited energy ($\sim$1/6) going into the He zone.  Since most of the energy is deposited in the H zones and mainly released by heating of the gas, it is expected that nebular phase spectra should be dominated by thermal line emission from H-rich material. The main cooling lines in the H zones during the first year are lines at UV wavelengths such as the \Mgii\ 2795, 2802 \AA, and \Feii\ 2382, 2600 \AA\ lines. However, there is also an appreciable contribution from cooling lines in the optical that can be analysed. Although the He zone gets $\sim$1/6 of the energy, it does not produce any distinct optical lines that can be studied. For most of the zones at 400 d, it can be seen that heating is the dominant channel for the deposited energy, with ionisation the next most important, while only a small percentage of energy goes into excitations.

The excitation of the lines can in general be divided into those excited thermally and those excited by recombinations.  
The recombination lines arise as a result of the non-thermal ionizations and in some cases, like H, photoionisations. They are therefore insensitive to the temperature, and mainly depend on the efficiencies for ionisation by the gamma-rays and the ionising UV radiation.
The thermally excited lines are responsible for most of the cooling of the ejecta. The strength of a given line from a specific abundance zone is determined by the fraction of energy deposited in that zone and the relative abundances in this zone. However, the efficiencies of different elements in cooling are very different, depending on their atomic structure. As an example the \Caii\ 7291, 7323 \AA\ lines are typically a factor of $10^2-10^3$ times more efficient per atom in cooling than the \Oi\ 6300, 6364 \AA\ lines. A small abundance of Ca can therefore dominate that of the much more abundant O \citep{fra89}. 
The luminosity, L($\lambda$) of the different lines can be estimated from Table \ref{model_tab} using, 
\begin{equation}
L(\lambda) = E_{tot} ~F_{i}~f_{heat}^{i}~f_{\lambda}^{i}
\end{equation}
where E$_{tot}$ is the total energy deposition (in the whole ejecta) and the other quantities are defined above and in Table \ref{model_tab}.
We discuss the individual lines here, referring specifically to the model described above.

\begin{table*}
\caption{Output values for the model at 400 d: total energy deposition, energy channels, and dominant cooling lines for each zone.}
\label{model_tab}
\begin{tabular}{llrrrrrrrrrrrrrrrrrrrrr }
\hline
\hline
\textbf{Zone} && \textbf{Fe/He} & \textbf{Si/S} & \textbf{O/Si/S}$^b$ & \textbf{O/Ne/Mg} & \textbf{O/C}$^b$ & \textbf{He}$^a$ & \textbf{H}$^a$ \\ \hline
Energy deposition \% (F$_{zone}$) & &7.5& 1.9& 2.5 & 2.5& 2.9&  16.2& 67\phantom{.0}\\ 
\hline
\textit{Breakdown of Energy Channels:}\\
Heating \%$^{a}$ ($f^{zone}_{heat}$)&& 76\phantom{.0}  & 52\phantom{.0}  & 43\phantom{.0}  & 52\phantom{.0}  & 43\phantom{.0}  & 43\phantom{.0}& 35\phantom{.0}\\
Ionisation \% ($f^{zone}_{ion}$)&&18\phantom{.0} & 29\phantom{.0}  & 47\phantom{.0}  & 42\phantom{.0}  & 49\phantom{.0}  & 44\phantom{.0}&37\phantom{.0} \\
Excitation \% ($f^{zone}_{exc}$) && 6\phantom{.0} & 19\phantom{.0}  & 10\phantom{.0} & 6\phantom{.0}  & 8\phantom{.0}  & 13\phantom{.0}& 28\phantom{.0}\\
\hline
\textit{Breakdown of Cooling Lines:} \\
\Caii\ (7291, 7323 \AA) ($f^{zone}_{7291}$)&&2.4&47\phantom{.0} &0\phantom{.0}&7.2&0\phantom{.0}&2.3&27\phantom{.0}\\
\Oi\ (6300, 6364 \AA) ($f^{zone}_{6300}$)&&0\phantom{.0} &0\phantom{.0} &0\phantom{.0}&55\phantom{.0}&0\phantom{.0}&0\phantom{.0}&4.4\\
\FeiiF\ (7155, 7172 \AA) ($f^{zone}_{7155}$)&&7.4&0.6&0\phantom{.0}&0.3&0\phantom{.0}&0.3&1.7\\
\hline
\hline
\end{tabular}\\
$^{a}$Energy deposition channels are calculated for the H and He zones using a mean, weighted by the energy deposition in each zone.\\
$^{b}$The zero values are due to molecule formation, which we assume is responsible for the cooling in these zones. \\
\end{table*}

\subsubsection{\ha} \label{ha_ha}

\ha\ is not an efficient cooler because the n=3 state has too high an excitation energy (10.2 eV) and  \ha\ is instead created by recombinations, and to a lesser extent by non-thermal excitations. 
The importance of photoionisations from the n=2 state by UV photons was first shown by \cite{kir75}.
The model results for the relative contributions are shown in Figure \ref{frac_hant} and found to be in good agreement with the results of \cite{xu92}. $f_{ion}^{H,nt}$ is the fraction of the energy deposited in the H zones going into non-thermal ionisations and $f_{ion}^{H,ph}$ is the fraction of the energy deposited in the H zones going into Balmer photoionisations. 
The value 0.14 is the fraction of non-thermal ionisation energy reemitted as \ha\ (1.89 eV/13.6 eV) and 0.56 is the fraction of Balmer photoionisation energy reemitted as \ha\ (1.89 eV/3.4 eV), where 1.89 eV is the energy of \ha\ photon and 3.4 eV is the ionisation potential of the Balmer continuum. 
Using our calculation, which includes a detailed computation of the internal UV field, we confirm the results in \cite{xu92} and \cite{koz92}, that Balmer photoionisations are dominant in the H ionisation rates up to $\sim$400--500 d. At this time a transition to the non-thermal dominated phase occurs, when the Balmer continuum becomes optically thin and direct ionisations from non-thermal electrons dominate. At 400 d, we find a deposition-weighted ratio of Balmer to non-thermal ionisation of $\sim$3 (see Figure \ref{frac_hant}).

\begin{figure}
\includegraphics[width=8.5cm]{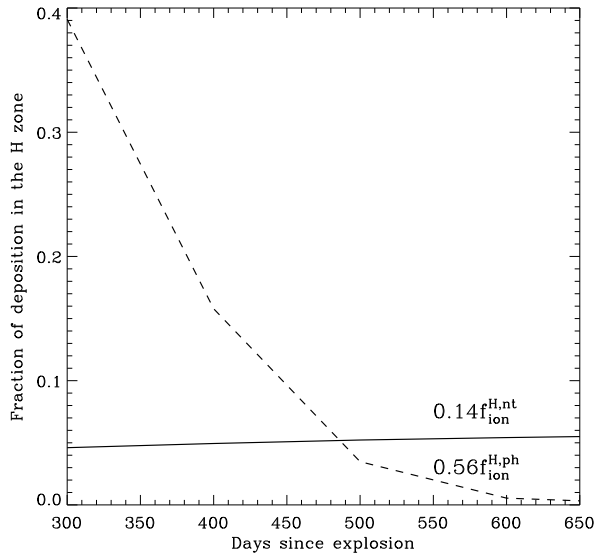} 
\caption{The contribution to the \ha\ luminosity from non-thermal ionisations, $f_{ion}^{H,nt}$ (solid line) and from photoionisations, $f_{ion}^{H,ph}$ (dashed line). }
\label{frac_hant}
\end{figure}

\begin{figure*}
\includegraphics[width=12cm]{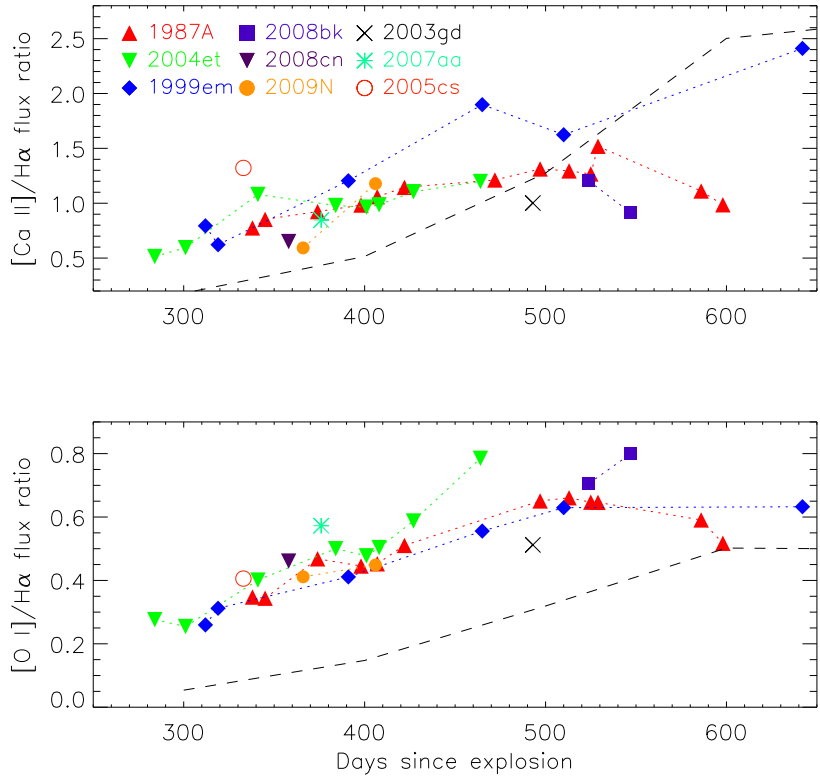}  
\caption{From top to bottom, the ratio of the flux in the \Caii\ doublet to \ha\ and the ratio of the \Oi\ doublet to \ha. The dashed black line shows the output from the model.}
\label{fluxratioocaha}
\end{figure*}

\subsubsection{\Oi\ 6300, 6364 \AA} \label{Oxygen}

The emission in the \Oi\ 6300, 6364 \AA\ lines is caused by a combination of emission from synthesised and primordial O. There are three zones that contain O as their main component; the O/Si/S, O/Ne/Mg, and the O/C zones. In the O/Si/S and O/C zones, modelling has shown that  molecules of SiO and CO form and are responsible for nearly all of the cooling \citep{liu95}. This is supported by the detection of SiO and CO in SN 1987A \citep{spy88, ait88}, as well as in other IIP SNe such as SN 2004et \citep{kot09, mag10}. Therefore, we can expect to see thermal emission from synthesised O from the O/Ne/Mg zone only. In this zone, the \Oi\ 6300, 6364 \AA\ doublet is one of the dominant cooling lines doing about half of the total cooling and provides a significant contribution to the \Oi\ 6300, 6364 \AA\ lines luminosity.

However, there is also a contribution to the \Oi\ doublet from primordial O in the H zones, where the doublet is found to do a few per cent of the total cooling. From Table \ref{model_tab} the fraction of the total emission in the \Oi\ lines from the O/Ne/Mg zone is $\sim 0.025 \times 0.52 \times 0.55 \approx 0.0072$, while it is $\sim 0.67 \times 0.35 \times 0.044 \approx 0.010$ in the H envelope. The low fraction of the cooling done by the \Oi\ lines in the H envelope is therefore balanced by the large total energy deposition in this zone. 

This leads to the important conclusion that in IIP SNe with progenitor masses of $\sim$12 \msun, the \Oi\ emission is formed by approximately equal contributions from primordial and synthesised O. This is perhaps surprising since there is $\sim$6 times more synthesised than primordial O present (assuming solar metallicity), but this is negated by the fact that the H zones absorb $\sim$8 times more gamma ray energy than the O zones, and also that the O/Si/S and O/C zones are cooled by molecules. 

This is different to the results that were obtained for SN 1987A using an 20 \msun\ model, where it was found that \Oi\ emission lines are formed in the zones containing synthesised O \citep{koz98}. \cite{woo95} showed that the O mass depends sensitively on the progenitor mass; a 20 \msun\ main-sequence star would produced an O mass of 1.8 \msun\ and the synthesised O mass would be $\sim$40 times greater than the primordial O mass, and so the contribution of synthesised O to the total \Oi\ 6300, 6364 \AA\ lines should dominate. These models for SN 1987A also assumed a lower amount of O in the envelope due to CNO mixing than obtained in the stellar evolution models used here.

\subsubsection{\Caii\ 7291, 7323 \AA}
Synthesised Ca exists mainly in the Si/S zone (2.1 $\times$ 10$^{-3}$ \msun), but also in the Fe/He zone (3.2 $\times$ 10$^{-4}$ \msun), while the total mass of primordial Ca present in the envelope is 6.4 $\times$ 10$^{-4}$ \msun ($\sim$25 per cent of the synthesised mass). As was previously shown by \cite{li93} and \cite{koz98} for SN 1987A, it is found for our 12 \msun\ model that the \Caii\ 7291, 7323 \AA\ doublet is formed nearly completely by primordial Ca in the inner regions of the H envelope. 
The \Caii\ 7291, 7323 \AA\ emission from the Si/S and Fe/He zones is no more than $\sim$10 per cent of the total flux at any time. This is because the masses of these zones are small compared to the H zones and so receive only a small amount of gamma-ray energy. At later times, the \Caii\ lines will receive important contributions from fluorescence \citep{li93} but at $\sim$400 d the \Caii\ emission from the H envelope is dominantly thermal.

\begin{figure*}
\includegraphics[width=12cm]{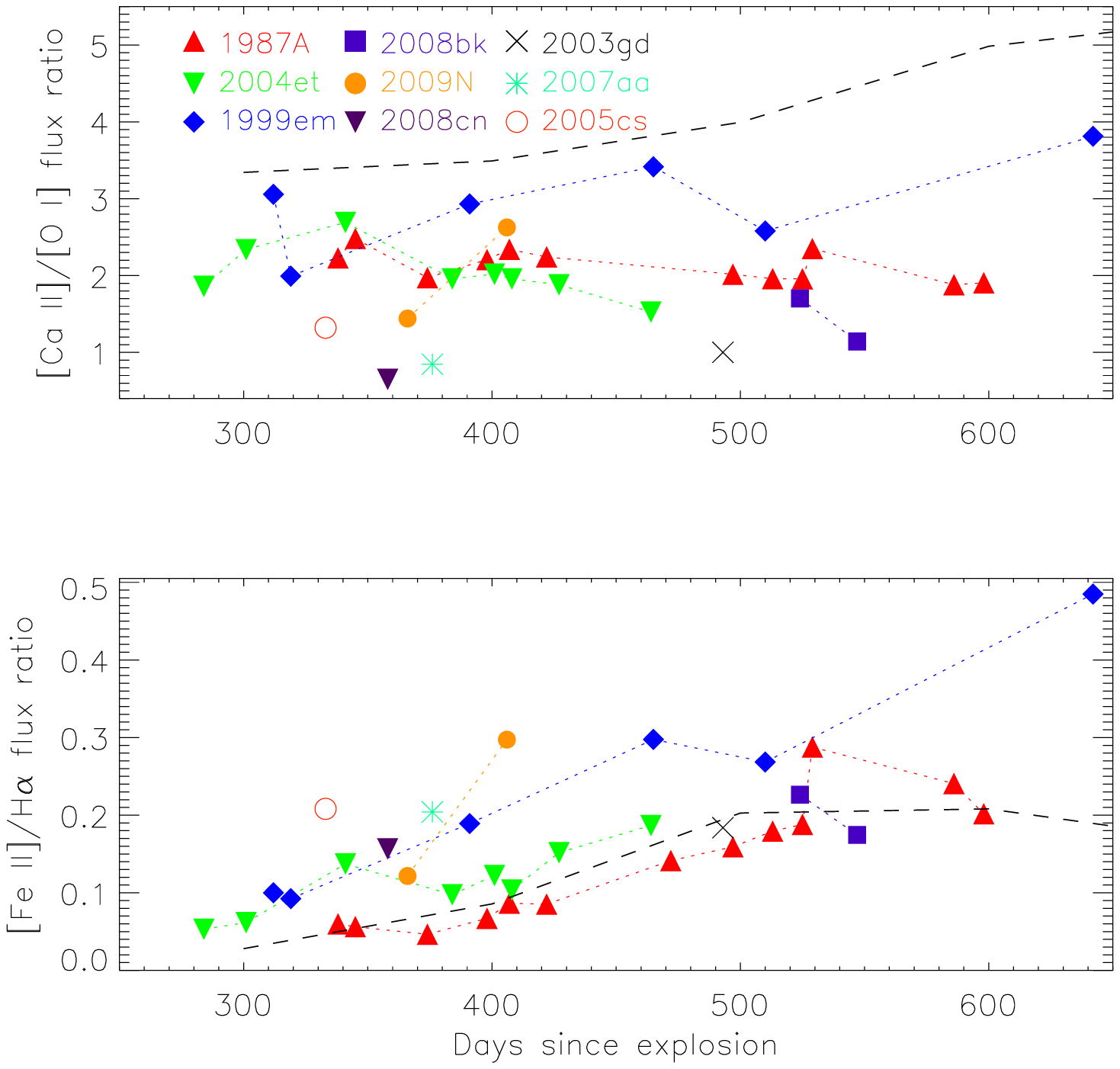}  
\caption{Ratio of flux in \Caii\ to \Oi\ and ratio of flux in \FeiiF\ to \ha. The dashed black line shows the output from the model.}
\label{fluxratiofe1}
\end{figure*}

\begin{figure*}
\includegraphics[width=12cm]{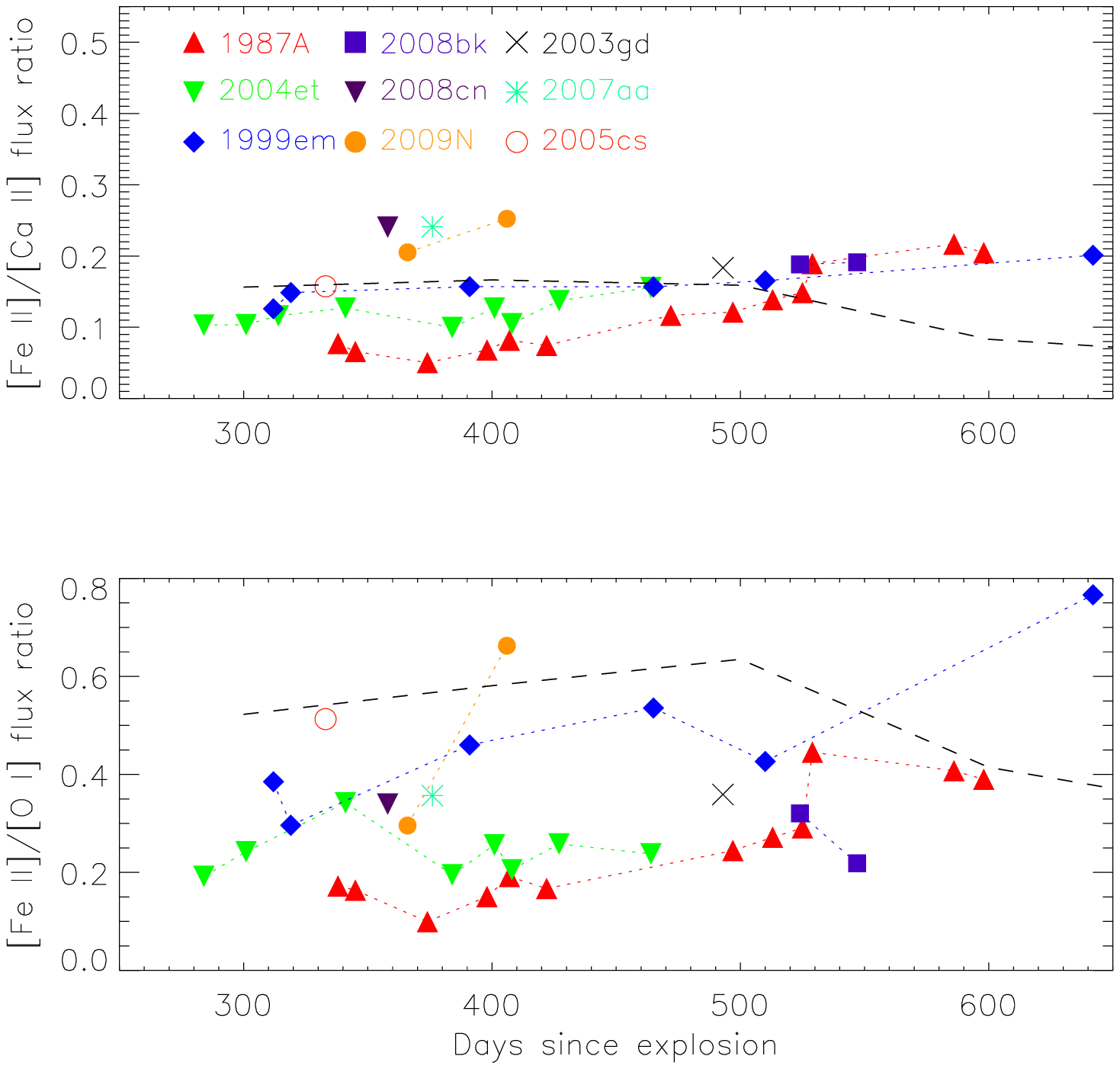}
\caption{Ratio of flux in  \FeiiF\  to \Caii\ and \Oi. The dashed black line shows the output from the model.}
\label{fluxratiofe2}
\end{figure*}

\subsubsection{\FeiiF\ 7155, 7172 \AA} \label{iron}
The \FeiiF\ 7155, 7172 \AA\ doublet has a flux ratio of its components of $\sim$3 to 1 if the parent multiplet is in local thermodynamic equilibrium and optically thin. The state from which the doublet comes is only 1.95 eV above the ground state, so this transition is thermally driven. The \FeiiF\ 7155, 7172 \AA\ doublet is formed by contributions from both synthesised Fe from the Fe/He zone and primordial Fe in the H envelope. 
From Table \ref{model_tab}, one finds from the products of the energy deposition, heating fraction and cooling fraction, that there are equal contributions from the Fe/He zone and the H envelope to this line.
This line, like the \Oi\  and  \Caii\ doublets, is therefore a mix of synthesised and primordial Fe, which is agreement with the results for SN 1987A \cite[e.g.][]{koz98}. 

\subsection{Application to observed line flux ratios} \label{lineratios}

Having obtained a theoretical interpretation of the origin of the flux in the prominent emission lines and the dominant processes involved (ionisation in the case of \ha\ and heating for the rest of the studied lines), these results can then be compared to the values obtained from our late-time nebular phase spectra using flux ratios of the prominent emission lines. Line ratios are a convenient method of comparing emission line properties of SNe both as a function of time and also between different objects, because any uncertainty in calibrating the absolute flux of the spectra is removed. 

The line ratios were measured by fitting a Gaussian to the line profiles and then integrating the flux under the curve and subtracting off a linear continuum. In the case of the doublets of \Oi, \Caii, and \FeiiF\ where the individual component lines were blended, both lines were fitted simultaneously with two Gaussians to obtain the total flux. In Figures \ref{fluxratioocaha}, \ref{fluxratiofe1} and \ref{fluxratiofe2}, the ratios of the prominent emission lines (\ha, \Oi\ 6300, 6364 \AA, \Caii\ 7291, 7323 \AA, and \FeiiF\ 7155, 7172 \AA) for the nine SNe described in Section \ref{SN_sample} are shown, along with the model curve obtained from the spectral synthesis model detailed above. The flux ratios outputed from the model are for a 12 \msun\ model with a core velocity of $\sim$1800 \kms. As noted in Section \ref{model_setup}, this model is not suitable for calculating line-flux ratios of low-luminosity, low-velocity events such as SN 2008bk (core velocity of $\sim$600 \kms) due to strong adiabatic cooling at these velocities, which makes steady-state temperature calculations unreliable. Therefore, comparisons of low-velocity events such as SNe 2008bk and 2009N with the model output in Figures \ref{fluxratioocaha}, \ref{fluxratiofe1} and \ref{fluxratiofe2} should be treated with caution.

Firstly, it can be seen in Figures \ref{fluxratioocaha} and \ref{fluxratiofe1} that there is a trend of increasing flux from the \Caii, \Oi, and \FeiiF\ lines with time compared to \ha. This can be understood based on the transition, as detailed in Section \ref{ha_ha}  of the Balmer continuum to an optically thin regime and shown in Figure \ref{frac_hant}. The contribution of the Balmer-continuum photoionisation to the total ionisation rate decreases with time and hence, the \ha\ luminosity also decreases. Also, we see in Figure \ref{model_eqn_comp} that the fraction of the total luminosity that is carried by the optical cooling lines increases up to 500--600 d, as the temperature falls to favour optical over UV line cooling. At $\sim$600 d, infrared lines take over and the optical lines decrease.

The \Caii\ 7291, 7323 \AA\ to \ha\ model ratio as a function of time is plotted in Figure \ref{fluxratioocaha} and found to be slightly lower at early times, but is in reasonable agreement with the observed flux ratios.
The \Caii\ to \ha\ ratio displays little variation between the different SNe and this can be understood by both of the lines being formed in the H zones, and therefore is independent of its mass, and the relative amounts of energy going into heating (responsible for the \Caii\ emission), and ionisation (responsible for the \ha\ emission) are not very sensitive to the density and temperature.

The \Oi\ 6300, 6364 \AA\ to \ha\ model ratio 
shows the same trend as the observed flux but is lower at all epochs. Since this is the same underproduction trend as seen at earlier times for the \Caii\ to \ha\ ratio, \ha\ may be overproduced in the model. The scatter in the ratio between different SNe does not seem to show any trend with \nick\ or progenitor mass. Although the synthesised O to H ratio should grow with progenitor mass, the significant contribution from primordial O will damp the impact of this on the \Oi\ 6300, 6364 lines. It is surprising that SN 1987A does not have the highest value since its \Oi\ emission was shown by \cite{koz98} to be dominated by emission from the O-rich zones, which should be significantly greater in a 20 \msun\ star compared to a low mass star. This suggests that either the mass of synthesised O
is smaller than previously believed, or that  SN 1987A is unusually H-rich.
 \begin{figure}
\includegraphics[width=8.5cm]{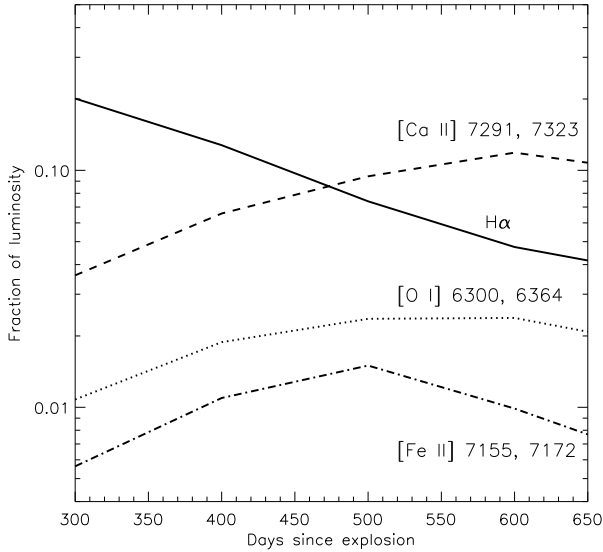} 
\caption{The fraction of the total luminosity emerging in \ha\ (solid line), \Oi\ 6300, 6364 \AA\ (dotted line), \Caii\ 7291, 7323 \AA\ (dashed line), and \FeiiF\ 7155, 7172 \AA\ (dash-dot line) in the spectral synthesis model.}
\label{model_eqn_comp}
\end{figure}

The \Caii\ to \Oi\ ratio
is also shown in Figure \ref{fluxratiofe1} and suggests that either \Caii\ is being overproduced in the model or \Oi\ underproduced. A wide range of observed values is seen for this ratio but there is no very noticeable trend with SN properties. There may be a weaker \Caii\ to \Oi\ flux ratio for lower velocity SNe such as SNe 2008bk and 2003gd seen at $\sim$500--550 d compared with the model, which could be explained by \Caii\ emission doing less cooling in the H zones at the lower temperatures (of the order of 1000K) seen in low \nick\ mass events (Jerkstrand et al.,~in prep.), but synthesised \Oi\ still doing most of the cooling in the O zones due to fewer competing cooling lines. However, this is a speculative explanation and the high-velocity SN 2008cn and `normal' velocity SN 2007aa also show weak \Caii\ to \Oi\ ratios at $\sim$350 d which cannot obviously be explained in this way. Further modelling using a variety of velocity and \nick\ mass inputs will be used to explore this in more detail.

The other prominent emission line studied here is the \FeiiF\ 7155, 7172 \AA\ doublet, which is detailed in \ref{iron}. The ratio of \FeiiF\ to \ha\ 
model curve is shown in Figure \ref{fluxratiofe1}, in good agreement with the observed range. There is a trend of increasing flux with time of \FeiiF\ with respect to \ha\ but no obvious trend between SNe with different \nick\ masses and velocities is seen.
The ratio of \FeiiF\ 7155, 7172 \AA\ doublet to the \Caii\ 7291, 7323 \AA\ line 
is plotted in Figure \ref{fluxratiofe2}, again good agreement between the model and observed ratios is seen. The \FeiiF\ to \Caii\ ratio seems to be relatively constant as a function of time and no obvious trend with \nick\ mass or velocity can be observed. 
The ratio of the \FeiiF\ 7155, 7172 \AA\ doublet to the \Oi\ 6300, 6364 \AA\ doublet 
can be formed using combinations of the primordial and synthesised O and Fe components,
and is shown in Figure \ref{fluxratiofe2}. The scatter in the observed values for the different SNe is quite large
 compared to the \FeiiF\ to \Caii\ ratio shown above, perhaps due to varying contributions from synthesised O to the \Oi\ lines.  

These comparisons are the first time that nebular 
phase spectra have been compared in detail and 
interpreted in the framework of a radiative transfer model. 
SN1987A has been the cornerstone of understanding 
synthesised material ejected in SNe in relation to a progenitor of
known mass, but we now have a sample of SNe with progenitor 
constraints and good quality nebular phase spectral coverage. No significant trends with progenitor mass are seen, with all the SNe in the sample having very similar flux ratios for the different elements. This suggests that we are in general, seeing emission from similar regions for the different SNe in the sample. The observed line widths and line profiles, along with the theoretical interpretation of the line ratios, suggest like for SN 1987A \citep{koz98, li93, li93b}, that \Caii\ emission is formed in the H zones and the \FeiiF\ emission is produced by a mixture of primordial and synthesised Fe.  

However, the most important result discovered here using the line profiles and widths and also the theoretical interpretation, is that \emph{for our 12 \msun\ model, the \Oi\ 6300, 6364 \AA\ emission lines are formed from nearly equal contributions from primordial O in the H-rich zones and synthesised O in the O-rich zones.} Figure \ref{prim_frac} shows a plot of the relative contribution from primordial abundances as a function of time for the \Oi, \Caii, and \FeiiF\ lines. At earlier times, the \Caii\ lines are mainly formed from thermal
processes, while at later times, there is also a contribution from
fluorescence. Both thermal emission and fluorescence occur mainly in the H
zone at all times. For the \Oi\ doublet, the division between synthesised and
primordial material is  $\sim$50 per cent each at all times. With time, the \FeiiF\ 7155, 7172 \AA\ doublet
becomes steadily more dominated by the H zone, as the Fe/He zone
rapidly cools to temperatures where infrared cooling dominates \citep{koz98b}. This is markedly different to the results obtained for SN 1987A, which showed that the emission was being produced nearly completely in the O-rich zones \citep{koz98, li92, spy91}. The reason for this is twofold; the
mass of synthesised O is $\sim$10 times larger in a 20 \msun\ star compared to 
a 12 \msun\ star, and the abundance of O in the envelope of SN 1987A was found to be strongly suppressed
by CNO mixing.

\subsection{Implications for synthesised O mass estimates}
In the past, the \Oi\ 6300, 6364 \AA\ flux has been used to estimate the mass of synthesised O and hence, the main sequence mass of the progenitor star. \cite{elm03a} used an equation that calculated the O mass in SN 1999em by using a ratio with SN 1987A involving its O luminosity in the \Oi\ 6300, 6364 \AA\ lines and its \nick\ mass during the tail phase. \cite{uom86} estimated the minimum mass of O needed to produce the \Oi\ emission lines during the nebular phase, using an equation that depended on the flux of the \Oi\ lines, the distance to the SN and the temperature of the emitting gas. The O mass of SN 1987A has been modelled in detail and found to be in the range 1.2--1.5 \msun\ \citep{koz98, chu94}. 

\begin{figure}
\includegraphics[width=8.5cm]{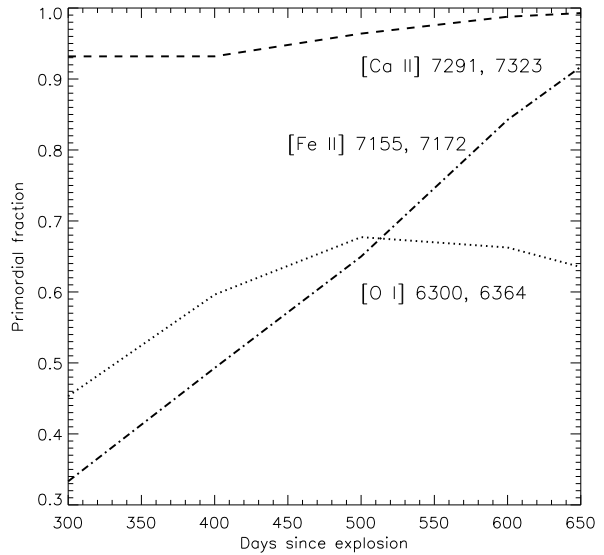}
\caption{Fraction of \Oi\ 6300, 6364 \AA\ (dotted line), \Caii\ 7291, 7323 \AA\ (dashed line), and \FeiiF\ 7155, 7172 \AA\ (dash-dot line) lines being produced by primordial abundances in the H zones relative to their total flux as a function of time since explosion.}
\label{prim_frac}
\end{figure}
However, unlike SN 1987A, the IIP SNe studied here result from the collapse of significantly less massive stars and we see evidence that the \Oi\ emission is made up of contributions from primordial and synthesised O, and so the \Oi\ doublet is not a simple, direct indicator of the synthesised O mass. Therefore, the O mass estimates in the literature for IIP SNe with low progenitor mass estimates should be treated with caution \citep[e.g.][]{and11, mag10, poz06, elm03a, elm11}, since they are based on an analysis of SN 1987A, and the true O masses could be only $\sim$50 per cent of these values. However, the O mass can still be estimated as long as the flux contribution from primordial O in the H zones is subtracted off first. An updated estimate of the synthesised O mass of SN 2004et is determined in Jerkstrand et al.~(in prep.) and can be used in the future for comparisons with SNe with low mass progenitor estimates.

\begin{figure}
\includegraphics[width=9.cm]{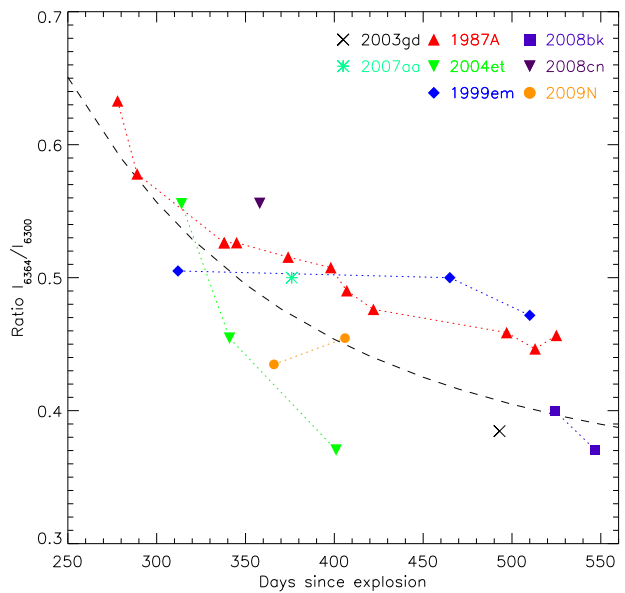}
\caption{Ratio of the flux in the [O I] 6364 \AA\ to the 6300 \AA\ lines. The black dashed line is the theoretical curve for an O number density of $7\times\ 10^{10}$ cm$^{-3}$ at 100 d.}
\label{ratioOx}
\end{figure}

\subsection{\Oi\ 6300, 6364 \AA\ line ratio}
The ratio of the 6364 \AA\ to the 6300 \AA\ line in the \Oi\ doublet can be used to determine whether the O-emitting zones are optically thick or optically thin \citep{spy91,li92}. In Figure \ref{ratioOx}, the ratio of the 6364 \AA\ to the 6300 \AA\ line is shown for a sample of eight IIP SNe. In the optically thick limit, the ratio of their intensities is 1:1, while in the thin limit the ratio of the intensities of the lines should be 1:3. An evolution from optically thick toward optically thin is seen for the sample of SNe in Figure \ref{ratioOx}. The theoretical curve for an O number density of $7\times\ 10^{10}$ cm$^{-3}$ at 100 d is also plotted in Figure \ref{ratioOx}. 

A contribution of up to $\sim$20 per cent from the \Fei\ 6361 \AA\ line to the flux in the \Oi\ 6364 \AA\ line was found for SN 1987A by \cite{koz98}. This was investigated for this model, but at 400 d, the \Fei\ 6361 \AA\ line was found to be a less than 3 per cent contaminant and no other lines were found in this wavelength region. Instead, we find that opacity due to other lines may influence the \Oi\ line ratio. The escape probability for photons, with respect to absorption in other lines is dependent on wavelength. For the two \Oi\ 6300, 6364 \AA\ lines there is a $\sim$60 per cent and $\sim$75 per cent chance of escaping, respectively. This will increase the 6364/6300 ratio, since the 6364 \AA\ photons will escape somewhat more easily (by a factor of 1.25 in this case).

\begin{figure*}
\includegraphics[width=15cm]{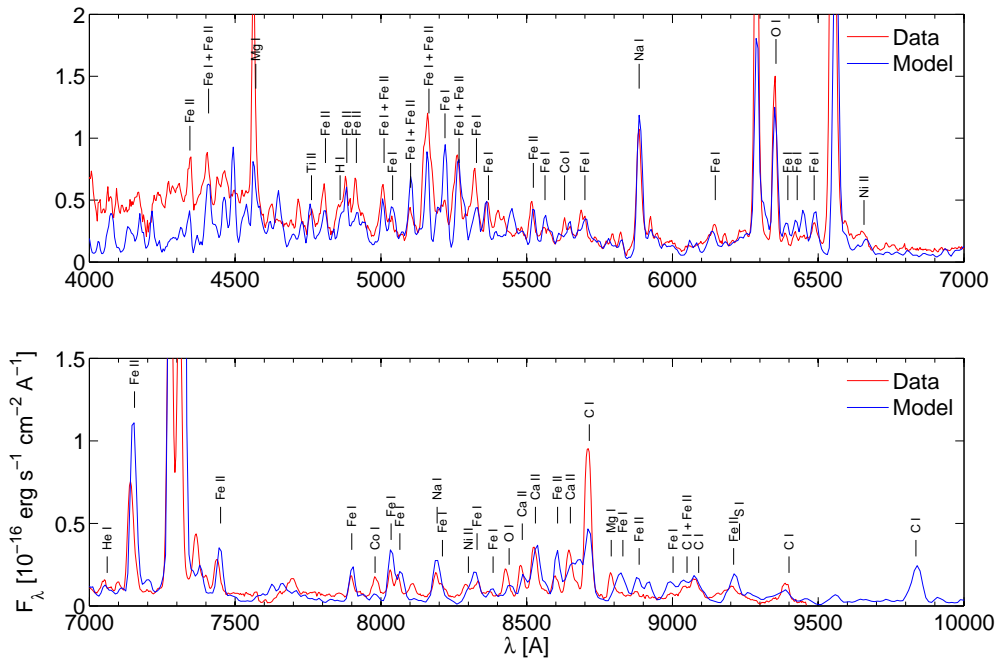} \caption{Line identification for the low-velocity, SN 2008bk at +547 d post-explosion comparing the spectral synthesis model (blue) to the data (red). We have not distinguished between forbidden and allowed lines, instead we show which ion contributes most to which feature.}
\label{08bkid}
\end{figure*}

There is a significant difference in the rate at which the O-emitting regions become optically thin for different SNe. SN 2004et shows a steep decline towards the optically thin regime, which might suggest that the density of its ejecta is low, which can be explained by the high HWHM \Oi\ velocities seen for SN 2004et of $\sim$1200 \kms\ during the nebular phases studied. In a similar way, the low velocity SN 2008bk had lower HWHM \Oi\ velocities ($\sim$350 \kms) and therefore remained denser for longer and hence took longer for its ejecta to become optically thin. However, SN 1987A like SN 2004et, had high \Oi\ velocities ($\sim$1100 \kms) and therefore, should have become optically thin at a similar rate to SN 2004et, but this evolution is not seen. This could be due to the large contribution from the synthesised \Oi\ in the core, where the densities are higher. 

Indeed, this ratio can be used to probe the density structure of the O-emitting regions as was done for SN 1987A by \cite{li92} and \cite{spy91}. However, unlike SN 1987A, it has been shown for lower mass stars, that the \Oi\ 6300, 6364 \AA\ doublet is formed equally by contributions from the envelope and O-rich zones and so any density estimates must use a two component model to fit the contributions from both the H-rich and O-rich zones.

\section{Line identifications of SN 2008bk} \label{line_id}

A late-time nebular spectrum (+547 d) of SN 2008bk is shown in Figure \ref{08bkid}, along with a model spectrum outputted from the spectral synthesis code. The identifications do not distinguish between forbidden and allowed lines, instead it identifies which element contributes most to which features. This SN has very low line velocities that gives narrow emission lines that are well resolved. In higher velocity events, such as SN 2004et, many lines are blended not due to the resolution limit of the instruments used to obtain the spectra but because their profiles intrinsically overlap in velocity space. Therefore, low velocity SNe give us a great opportunity to identify individual line emissions that would be blended in higher velocity events. The prominent emission lines in this spectrum were identified using the same spectral synthesis code as discussed earlier, again using a 12 \msun\ model, but this time choosing a core velocity of 600 \kms\ to match the velocities seen in SN 2008bk. Adiabatic is strong relative to line cooling in the envelope, as the low \nick\ mass implies a low electron fraction and low collision frequencies. This makes the steady-state temperature calculation unreliable and therefore, this model has not been used to calculate flux ratios. However, the line identifications should not be affected and can be used to identify the spectral features in low velocity spectra such as those of SN 2008bk. 

Most of the lines in the complex region from 4000--5500 \AA\ can be identified
with \FeiF\ and \FeiiF\ lines, created by scattering and fluorescence in the H
envelope. 
Above 5500 \AA\, the core zones start making contributions to the spectrum.
The 5900 \AA\ line is \Nai\ 5896 \AA\, which scatters emission from
the \Nai\ 5890 \AA\ line, as well as from the \Hei\ 5876 \AA\ line, with about a third of the photons emitted by synthesised Na in the O/Ne/Mg zone.
The \Fei\ lines at 7900, 8030, 8060, 8200, 8300, and 8350 \AA\ are all from Fe
in the core (both from the Fe/He clumps and the O/Ne/Mg clumps).
In the near-infrared, we can identify the \Oinir\ 8446 \AA\ line, which is usually hard to
see due to blending with the \CaiinF\ triplet in higher-velocity SNe. Another line
which usually also suffers from this blending is \CiF\ 8727 \AA\, which is clearly
seen here. In the model, this line, along with the \Ci\ lines at 9000--9100 \AA\ and 9400 \AA\ are created by C in the He envelope, verifying that significant dredge-up has occurred in this star.

\section{Correlation of velocity with \nick\ mass}

There are currently a number of methods by which the \nick\ mass ejected in a IIP SN can be estimated. Most methods require knowledge of the bolometric light curve of the SN during the early radioactive tail phase. During this phase, the bolometric light curve is powered nearly completely by the radioactive decay of $^{56}$Co to $^{56}$Fe and hence the luminosity at this stage gives a good indication of the synthesised \nick\ mass that is ejected in the explosion. The link between the light curve and \nick\ mass can be made in three different ways: a comparison with the luminosity of SN 1987A, which has a well constrained \nick\ mass, the steepness parameter method of \cite{elm03b}, and the tail luminosity method of \cite{ham03}. However, the use of these methods is limited to SNe with good bolometric coverage during the radioactive tail phase, as well as reliable distance and extinction estimates.

For SNe with limited multi-band coverage, \cite{ber09} and \cite{mag10} described bolometric corrections to convert from a magnitude obtained in a single band to a bolometric luminosity. The former paper estimated the correction as a function of colour, while the latter work as a function of time since explosion. These corrections are useful for sparsely covered events, but still require an accurate estimate of the extinction and distance to the SN. Using late-time spectra (200--400 d post-explosion), \cite{chu90} and \cite{elm03b} described how the luminosity of the \ha\ emission line is proportional to the ejected \nick\ mass. This method is useful when no photometry is available, but again the distance and extinction to the SN must be known.  Therefore, independent methods that do not rely on knowledge of the absolute magnitude of the SN would be a useful addition. \cite{ham03} showed a correlation between the \nick\ mass and the photospheric velocity at mid-plateau. However, the photospheric velocity decreases rapidly during the photospheric phase and therefore, requires a well-sampled light curve to provide a secure constraint on the explosion epoch to obtain the velocity mid-plateau. In this section, we discuss a method for estimating the ejected \nick\ mass from the FWHM of the emission lines observed in nebular phase spectra of IIP SNe, where the velocity is found to relatively constant in the range $\sim$350-550 d post explosion and hence does not require accurate constraints on the explosion epoch.
 \begin{figure}
\includegraphics[width=8.5cm]{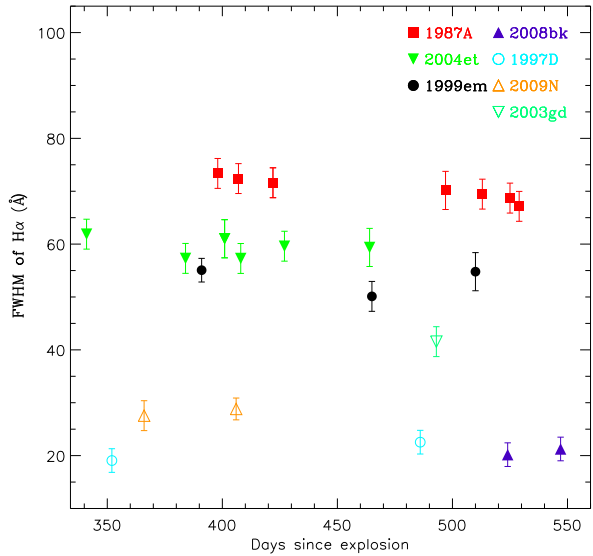} 
\caption{The FWHM$_{\rm corr}$ of the \ha\ emission line corrected for the instrumental resolution for SNe 1987A, 1997D, 1999em, 2003gd, 2004et, 2008bk, and 2009N as a function of time since explosion.}
\label{fwhm_allha}
\end{figure}

The sample of SN spectra used in this work is detailed in Section \ref{SN_sample}. The sample is a combination of IIP SNe found in the literature with good photometric and spectral coverage during the nebular phase, along with new previously unpublished observations. The photometric coverage was necessary so that the \nick\ mass could be estimated independently from the spectra and an empirical relation between the ejected \nick\ mass and the line widths could be determined. We have added one more IIP SNe with nebular phase observations and reliable \nick\ mass estimates to the sample, SN 1997D, which was a low-luminosity, low velocity SN with an ejected \nick\ mass of 0.008 \msun\ \citep{pas04}. SN 1997D is not studied in detail here because of the lack of information on its progenitor star. The spectra of SN 1997D are taken from \cite{ben01} with a spectral resolution of 13 \AA. 

The FWHM of the most prominent feature in the spectra, the \ha\ 6563 \AA\ emission line was measured by fitting a Gaussian to the emission profile. This line is very strong in all the spectra and is unblended with other lines so we can be confident that the FWHM can be measured to good accuracy. As in Section \ref{vel_dist}, the FWHM of the line profiles were corrected for the spectral resolution of the instruments used to obtain each spectrum.
Figure \ref{fwhm_allha} shows the FWHM$_{\rm corr}$ of the \ha\ emission line as a function of time for 8 IIP SNe that also have \nick\ masses estimated from their bolometric light curves. From Figure \ref{fwhm_allha}, it can be seen clearly that for any given SN, the FWHM$_{\rm corr}$ is relatively constant at these epochs and so observations at a coeval epoch are unnecessary for a comparison to be made. A mean value of the FWHM$_{\rm corr}$ was calculated for each SN from the available data at epochs in the range $\sim$350-550 d. These values can then be compared to the \nick\ mass estimated from the luminosity of the nebular phase light curve. The FWHM$_{\rm corr}$ value used for each SN along with its \nick\ mass estimate from the literature are given in Table \ref{fwhm_nimasstable}.

Figure \ref{fwhm_nimassha} shows the ejected \nick\ mass calculated from the tail phase luminosity of the SNe against the FWHM$_{\rm corr}$ of the \ha\ emission line. It is immediately apparent that a correlation exists between these parameters, with small FWHM$_{\rm corr}$ (low velocity) events such as SNe 1997D, 2008bk, and 2009N displaying correspondingly smaller \nick\ masses. In a similar way, the higher velocity SNe of the sample, such as SNe 1987A and 2004et, have higher \nick\ masses. The relationship between the mass of ejected \nick\ and the FWHM$_{\rm corr}$ of the \ha\ emission line can be fitted with an equation of the form,
\begin{figure}
\includegraphics[width=8.5cm]{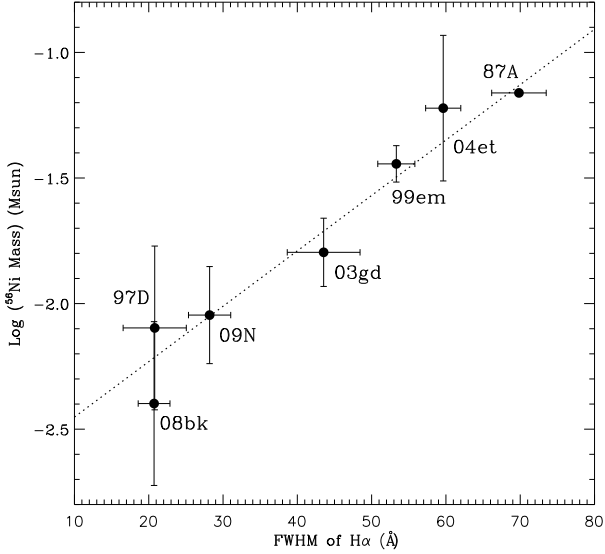}
\caption[The \nick\ mass obtained from the bolometric light curve against the FWHM of \ha.]{The \nick\ mass obtained from the tail phase of the bolometric light curve against the FWHM of the \ha\ emission line for the SNe listed in Table \ref{fwhm_nimasstable}.}
\label{fwhm_nimassha}
\end{figure}
 \begin{equation}
\label{ni_corr}
M({}^{56}{\rm Ni}) = A \times 10^{B ~\rm FWHM_{corr}} ~\msun
\end{equation}
where $B = 0.0233 \pm 0.0041$ and $A=1.81^{+1.05}_{-0.68}\times10^{-3}$. The reduced $\chi^2$ value for this fit is 0.73. This equation can used for SNe for which limited photometric data are available or where there are large uncertainties on the distance and extinction estimates. 

\begin{table}
\caption[Data used to determine the relationship between the ejected \nick\ mass and FWHM of \ha.]{Data used for Figure \protect  \ref{fwhm_nimassha} to determine the relationship between the ejected \nick\ mass and the FWHM of \ha\ corrected for the spectral resolution.}
\label{fwhm_nimasstable}
\begin{center}
\begin{tabular}{@{}lccl}
\hline
SN  & FWHM$_{\rm \ha}$ (\AA)& $^{56}$Ni mass (\msun)& Reference\\
\hline
1997D&20 $\pm$ 4&		0.008 $\pm$ 0.006 & \cite{ben01}\\
2008bk&21 $\pm$ 2	&	0.004 $\pm$ 0.003 &  Pignata et al.~(in prep.)\\
2009N&28 $\pm$ 3	&	0.009 $\pm$ 0.004& Tak\'ats et al.~(in prep.)\\
2003gd&44 $\pm$ 5&		0.016 $\pm$ 0.01\phantom{0} & \cite{hen05}\\
1999em&51 $\pm$ 2 &		0.036 $\pm$ 0.005 & \cite{utr07}\\
2004et& 60 $\pm$ 2&	0.064\phantom{0} $\pm$ 0.04\phantom{0} & \cite{mag10}\\
1987A&70 $\pm$ 4  &	0.069 $\pm$ 0.0012 &\cite{bou91}\\
\hline
\end{tabular}
 \end{center}
\end{table}

As an example we apply this relation to two IIP SNe, without good late-time light curves. SN 2008cn is a IIP SN that was observed spectroscopically at the VLT at an epoch of +358 d post-explosion. No photometric data has yet been published on this object, but from our spectrum, using a FWHM$_{\rm corr}$ of \ha\  of 79 \AA, we have estimated the \nick\ mass to be  0.13$_{-0.09}^{+0.30}$ \msun. SN 2007aa is another IIP SN that was observed at the VLT at a nebular phase epoch of +376 d post-explosion and found to have for \ha\ a FWHM$_{\rm corr}$ of 52 \AA\, which gives a \nick\ mass of 0.03$_{-0.02}^{+0.05}$ \msun. Hopefully these \nick\ mass estimates will be confirmed by future data releases for these SNe. The uncertainties in these estimates are relatively large, but at the very least this method can be used to determine between low, intermediate and high ejected \nick\ masses. A larger sample is needed to study this method of \nick\ mass determination in more detail and in particular it would be of interest to include some IIL SNe to see if the relation also holds for these events.

A correlation between the \nick\ mass and explosion energy for IIP SNe was shown by \cite{ham03} for a sample of 16 IIP SNe, with higher \nick\ mass events having greater explosion energies. \cite{nad03} also describe a correlation between the explosion energy and the \nick\ mass of IIP SNe, which was suggested to have a complex nature that also depends on the progentior mass. One can assume that the velocity (or equivalently the FWHM) of the line profiles will scale with the explosion energy (which is a combination of internal and kinetic energy), since the more energy produced in the explosion, the more kinetic energy that will be given to the ejecta (at least 90 per cent of the explosion energy \citep{arn96}). This is the first time a correlation between the FWHM of the nebular phase line profiles and the \nick\ mass has been identified, and assuming similar ejecta masses for the SN sample, can be interpreted as a correlation between the explosion energy and the \nick\ mass.

\section{Summary and conclusions}

In this paper we have investigated the nebular phase spectra of a
sample of IIP SNe and compared their observational analysis to a
detailed radiative transfer model, to try to better understand the
formation of their emission lines and the variation within 
the sample. We have focused on IIP SNe with either progenitor detections
or restrictive limits on their progenitor masses.  Despite the large variation in \nick\ masses
and expansion velocities, the characteristics of the emission lines in
the nebular spectra of IIP SNe are found to be relatively homogeneous and independent of progenitor masses. For each SN, the shapes of the different emission line profiles are very similar and suggest a similar spatial distribution of the different elements. This is also seen when the velocities of the prominent emission lines are measured using their HWHM, with only \ha\ showing a noticeably higher velocity, which suggests a larger contribution from farther out in the ejecta. 

For SN 1987A, as has been previously shown, we have shown that the \ha\ velocity was significantly larger than the \Oi\ velocity, while for the other SNe in our sample the differences between H and \Oi\ were not as pronounced. This suggests that the zones responsible for \Oi\ emission are different in the rest of the SN sample compared to those in SN 1987A. From analysis of the line profiles and widths, it appears that all IIP SNe (except SN 1987A) have similar spatial distributions of \Oi\ which is closer in velocity to \ha\ than for SN 1987A. This implies that the \Oi\ flux could be predominantly emitted by primordial O in the H zones and may not be coming from the synthesised O from the inner regions of the ejecta. The line profiles are also shown to be intrinsically peaked in shape, even after taking into account the convolution with the instrumental resolution. This shows that mixing must have occurred for all the elements to be producing emission from both the centre and the outer regions of the ejecta.

These observational results are corroborated by the output of the radiative transfer model of a 12 \msun\ star,  where a significant proportion of the flux of the prominent emission lines is found to originate in the H-rich zones.  For SN 1987A, the \Caii\ and several \FeiiF\ emission lines were also found to be predominantly formed in the H zones, but here we present the new result that for the 12 \msun\ progenitor model used here, the \Oi\ flux is produced in approximately equal amounts by O in the H zones and O in the inner O-rich zones, and this explains the much smaller difference between the \ha\ and \Oi\ velocities seen for our sample of SNe. This has major implications for current methods of calculating the synthesised O mass in IIP SNe, which assume that the \Oi\ 6300, 6364 \AA\ line emission is produced solely by synthesised O. A calculation of the synthesised O mass of SN 2004et, that takes into account these new results will be presented in Jerkstrand et al.~(in prep.). 

The flux ratios of the emission lines from the 12 \msun\ model were found in general to agree quite well with the observed flux ratios. No trends were found between progenitor mass and the observed flux ratios, in agreement with the result that the flux in these lines has a significant contribution from primordial abundances. We also investigated possible correlations between the spectral properties and other observational properties of the SNe and found a correlation between the ejected \nick\ mass and FWHM of the \ha\ emission line. An empirical relation was obtained, which can be used to estimate the \nick\ mass of SNe with limited bolometric coverage and/or poor constraints on their extinction and distance. The benefit of this method is that it requires only the width of the \ha\ emission line during the nebular phase to estimate the \nick\ mass. Even more importantly, this provides a strong constraint on explosion models, particularly the amount of fall-back material in them.

\section{Acknowledgements}
C.~F. and A.~J. are supported by the Swedish Research Council and the Swedish National Space Board. S.~B., F.~B.~and A.~P.~are partially supported by the PRIN-INAF 2009 with the project ``Supernova variety and nucleosynthesis yields". G.~L.~is supported by a grant from the Carlsberg foundation. The Dark Cosmology Centre is funded by the Danish National Research Foundation. This work made use of the 
code developed by \cite{koz92} for the computation of
the non-thermal electron degradation.
This work has made use of data taken at the VLT under programme numbers 080.D-0213(B), 084.D-0261(B), 083.D-0131(B) and 084.D-0261(B) and at the NTT under programme numbers 083.D-0970(A) and 184.D-1140(S).

\end{document}